\documentclass[9pt,conference]{IEEEtran}
\IEEEoverridecommandlockouts
\usepackage{cite}
\usepackage{amsmath,graphicx}
\usepackage{epsfig}
\usepackage{subfigure}
\usepackage{multirow}
\usepackage{hhline}
\usepackage{color}

\begin{document}

\title{Center-Extraction-Based Three Dimensional Nuclei Instance Segmentation of Fluorescence Microscopy Images
\thanks{This work was partially supported by a George M. O'Brien Award from the National Institutes of Health under grant NIH/NIDDK P30 DK079312 and the endowment of the Charles William Harrison Distinguished Professorship at Purdue University.}
}

%\author{David~Joon~Ho,
        %Shuo~Han,
				%Chichen~Fu,
				%Paul~Salama,
				%Kenneth~W.~Dunn,
        %and~Edward~J.~Delp%
%\thanks{David Joon Ho, Shuo Han, Chichen Fu, and Edward J. Delp are with the Video and Image Processing Laboratory, School of Electrical and Computer Engineering, Purdue University, West Lafayette, IN 47907 USA.}% <-this % stops a space
%\thanks{Paul Salama is with the Department of Electrical and Computer Engineering, Indiana University-Purdue University, Indianapolis, IN 46202 USA.}% <-this % stops a space
%\thanks{Kenneth W. Dunn is with Division of Nephrology, School of Medicine, Indiana University, Indianapolis, IN 46202 USA.}%
%}
\author{\IEEEauthorblockN{David Joon Ho}
\IEEEauthorblockA{\textit{Video and Image Processing Laboratory}\\
\textit{School of Electrical and}\\
\textit{Computer Engineering}\\
\textit{Purdue University}\\
West Lafayette, Indiana}
\and
\IEEEauthorblockN{Shuo Han}
\IEEEauthorblockA{\textit{Video and Image Processing Laboratory}\\
\textit{School of Electrical and}\\
\textit{Computer Engineering}\\
\textit{Purdue University}\\
West Lafayette, Indiana}
\and
\IEEEauthorblockN{Chichen Fu}
\IEEEauthorblockA{\textit{Video and Image Processing Laboratory}\\
\textit{School of Electrical and}\\
\textit{Computer Engineering}\\
\textit{Purdue University}\\
West Lafayette, Indiana}
\and
\IEEEauthorblockN{Paul Salama}
\IEEEauthorblockA{\textit{Department of Electrical and}\\
\textit{Computer Engineering}\\
\textit{Indiana University-Purdue University}\\
Indianapolis, Indiana}
\and
\IEEEauthorblockN{Kenneth W. Dunn}
\IEEEauthorblockA{\textit{Division of Nephrology}\\
\textit{School of Medicine}\\
\textit{Indiana University}\\
Indianapolis, Indiana}
\and
\IEEEauthorblockN{Edward J. Delp}
\IEEEauthorblockA{\textit{Video and Image Processing Laboratory}\\
\textit{School of Electrical and}\\
\textit{Computer Engineering}\\
\textit{Purdue University}\\
West Lafayette, Indiana}
}

\maketitle

\begin{abstract}
Fluorescence microscopy is an essential tool for the analysis of 3D subcellular structures in tissue. An important step in the  characterization of tissue involves nuclei segmentation. In this paper, a two-stage method for segmentation of nuclei using convolutional neural networks (CNNs) is described. In particular, since creating labeled volumes manually for training purposes is not practical due to the size and complexity of the 3D data sets, the paper describes a method for generating synthetic microscopy volumes based on a spatially constrained cycle-consistent adversarial network. The proposed method is tested on multiple real microscopy data sets and outperforms other commonly used segmentation techniques.
%Nuclei segmentation is an important step used to characterize the tissue.
%In this paper, a two-stage method for segmentation of nuclei using convolutional neural networks (CNNs) is described.
%The first CNN selects coordinates for nuclei centers and generates a binary segmentation mask whereas the second CNN segments each nucleus.
%Manual segmentation for creating labeled volumes for training is not practical due to the size and complexity of the 3D data sets.
%To train our CNNs, we generate synthetic microscopy volumes based on a spatially constrained cycle-consistent adversarial network.
%Our method is tested on multiple real microscopy data sets and outperforms the segmentation techniques tested.
\end{abstract}


\vspace{-0.05in}
\section{Introduction}
\vspace{-0.05in}
\label{sec:intro}

Optical fluorescence microscopy enables imaging three dimensional subcellular components in tissue \cite{dunn2002}.
In particular, two-photon microscopy allows imaging deeper into the tissue with near-infrared excitation light \cite{vonesch2006}.
Three dimensional segmentation of subcellular components, such as nuclei, is required to quantify and analyze the microscopy volumes.
It is tedious to manually create labeled ground truth volumes for training machine learning methods. Moreover, this task is further complicated when nuclei are touching.
%It is tedious for one to manually analyze large and complex volumes especially in 3D.
%This is particularly true for creating labeled ground truth volumes for training machine learning methods.
%Below we will overview recent methods for nuclei segmentation and describe our approach to segmentation.
%It is also challenging to segment nuclei because some nuclei are touching. % \cite{xing2016}.

Watershed techniques which select local maxima of a distance transform as markers have been used to segment touching nuclei \cite{vincent1991}.
In \cite{yang2006} watershed markers are selected based on mathematical morphology to segment nuclei in time-lapse microscopy.
Watershed approaches generally over-segment nuclei due to their irregular structures.
To circumvent this, deformable models such as active surfaces have been investigated \cite{delgado2015}.
A method using multiple active surfaces was introduced to separate touching nuclei wherein the energy functional includes a penalty term for overlapping nuclei and a constraint term for volume conservation \cite{dufour2005}.
%Alternatively, active mask segmentation combining multiresolution, multiscale, and region-growing techniques is described in \cite{srinivasa2009}.
Alternatively, a method, known as Squassh, couples image restoration and segmentation by using an energy functional  derived from a generalized linear model \cite{rizk2014}. %paul2013
A common issue that arises is that  these methods frequently cannot distinguish nuclei from other biological structures.

Recently, convolutional neural networks (CNNs), that rely on the availability of large amounts of labeled training images, have been used  for many computer vision problems \cite{lecun2015}.
%Recently, convolutional neural networks (CNNs) have been used  for many computer vision problems with the advancement in graphics processing units (GPUs) and the availability of large amounts of labeled training images \cite{lecun2015}.
CNNs have very much impacted biomedical image analysis \cite{litjens2017}. 
%The use of CNNs for biomedical image analysis has increased the issues related to the need for large amounts of ground truth data needed for training.
%We have previously demonstrated in \cite{fu2017} a CNN nuclei segmentation technique that distinguishes nuclei from other structures.
%We used a set of actual ground truth images to train our 2D CNN.
%Segmentation results in horizontal, frontal, and segittal planes using a CNN were combined by a majority voting approach for 3D nuclei segmentation.
A deep contour-aware-network is described for gland segmentation in \cite{chen2016}.
The network produces object segmented images and contour segmented images where the contour segmented images are used to separate touching glands.
In \cite{graham2018} weights are assigned to the boundary of nuclei in hematoxylin and eosin (H\&E) stained histology images during  training to ensure touching nuclei are separated.
%Alternatively, a CNN-based cell detection method using a star-convex polygon and predicting object probabilities and their radian distance in various orientations is used to separate objects in \cite{schmidt2018}.
More recently, a cell detection and segmentation technique is presented in \cite{falk2019} using a U-Net architecture \cite{cicek2016}.

One challenge of using CNNs in biomedical image analysis is the lack of labeled training images due to the expensive and tedious labeling process. % needed to ground truth biomedical images.
Data augmentation techniques using simple transformations can be used to generate more training images but they still require labeled training images.
%Data augmentation techniques can be used to generate more training images using simple transformations on a limited number of labeled training images and can avoid overfitting problems. % \cite{krizhevsky2012}.
%In some cases simple transformations do not generate adequate training data, which has caused interest in using more sophisticated approaches to generating synthetic images that can be used for training.
To address the problem of limited availability of 3D ground truth volumes, we described in \cite{ho2017b} the generation of 3D synthetic microscopy volumes without using any labeled  volumes. 
The synthetic volumes were generated using a statistical model and a simple model of the point spread function of the microscope with ellipsoidal shaped nuclei.
%The synthetic volumes were generated using a statistical model and a simple model of the point spread function of the microscope.
%We used this approach to generate volumes with ellipsoidal shaped nuclei.
The synthetic volumes are then used to train CNNs to segment nuclei in real microscopy volumes.
We also presented a 3D detection and segmentation method in \cite{ho2018} using synthetic microscopy volumes generated similar to our previous work described in \cite{ho2017b}.

There has been a great deal of work in generating  realistic synthetic images that can be used for training using generative adversarial networks (GANs) \cite{goodfellow2014}.
%In \cite{isola2017} an image-to-image translation is described where a GAN is trained by a set of paired images, one of the images is an actual ground truth image, to generate synthetic images. 
%To use this GAN approach one needs access to a small set of actual ground truth images which may not be possible in some applications.
A CycleGAN was introduced where a GAN with a cycle consistency term can produce synthetic images that can be used for training without access to any actual ground truth images \cite{zhu2017}.
We described a spatially constrained CycleGAN (SpCycleGAN) in \cite{fu2018} to generate synthetic images  where a spatial constraint term is included in the CycleGAN.
We then trained a CNN using the synthetic volumes generated by the SpCycleGAN to produce accurate binary segmentation masks \cite{fu2018}. 
One problem in \cite{fu2018} is that we could not distinctly label each nucleus accurately.

In this paper, we present a 3D nuclei instance segmentation method using two CNNs for fluorescence microscopy volumes.
We define ``instance segmentation'' as a process where each object is detected and segmented with distinct labels.
%We use synthetic microscopy volumes generated by our SpCycleGAN described above and in \cite{fu2018} to train the segmentation method.
%Note no actual ground truth volumes are used for generating the synthetic volumes.
%In this paper we extract the central area of nuclei where the central regions do not overlap with each other even when the surface of the nuclei may overlap.
%In this paper we do not extract the nuclei boundary explicitly but  we extract the central area of nuclei where the central regions do not overlap with each other even when the surface of the nuclei may overlap.
%Our method is composed of two stages where nuclei locations are detected in the first stage and individual nuclei are segmented in the second stage.
This paper is different from our work described in \cite{ho2018} that detects locations of nuclei using a distance transform causing over-detection of irregular nuclei structures and segments each nucleus using a CNN trained by a set of blurred and noisy synthetic volumes generating inaccurate segmentation masks.
In the present paper, we use realistic synthetic training volumes generated by the SpCycleGAN \cite{fu2018} to train one CNN to detect the location of nuclei and a second CNN to segment each nucleus accurately.
Note no actual ground truth volumes are used for generating the synthetic volumes.
During detection we extract the central area of nuclei that do not overlap with each other even when the surfaces of the nuclei may overlap.
%This paper is different from our 2018 work described in \cite{ho2018} that detects locations of nuclei using a distance transform and segments each nucleus using a CNN.
%The CNN in \cite{ho2018} was trained by a set of blurred and noisy synthetic volumes.
%In the present paper, we use realistic synthetic training volumes generated by the SpCycleGAN to accurately detect and segment nuclei for instance segmentation.
We evaluate our method using a ground truth volume generated from a real fluorescence microscopy volume from a rat kidney.
Our data are collected using two-photon microscopy where nuclei labeled with Hoechst 33342 stain.

\vspace{-0.05in}
\section{Proposed Method}
\vspace{-0.15in}
\label{sec:method}
\begin{figure}[htb]
\centerline{\epsfig{figure=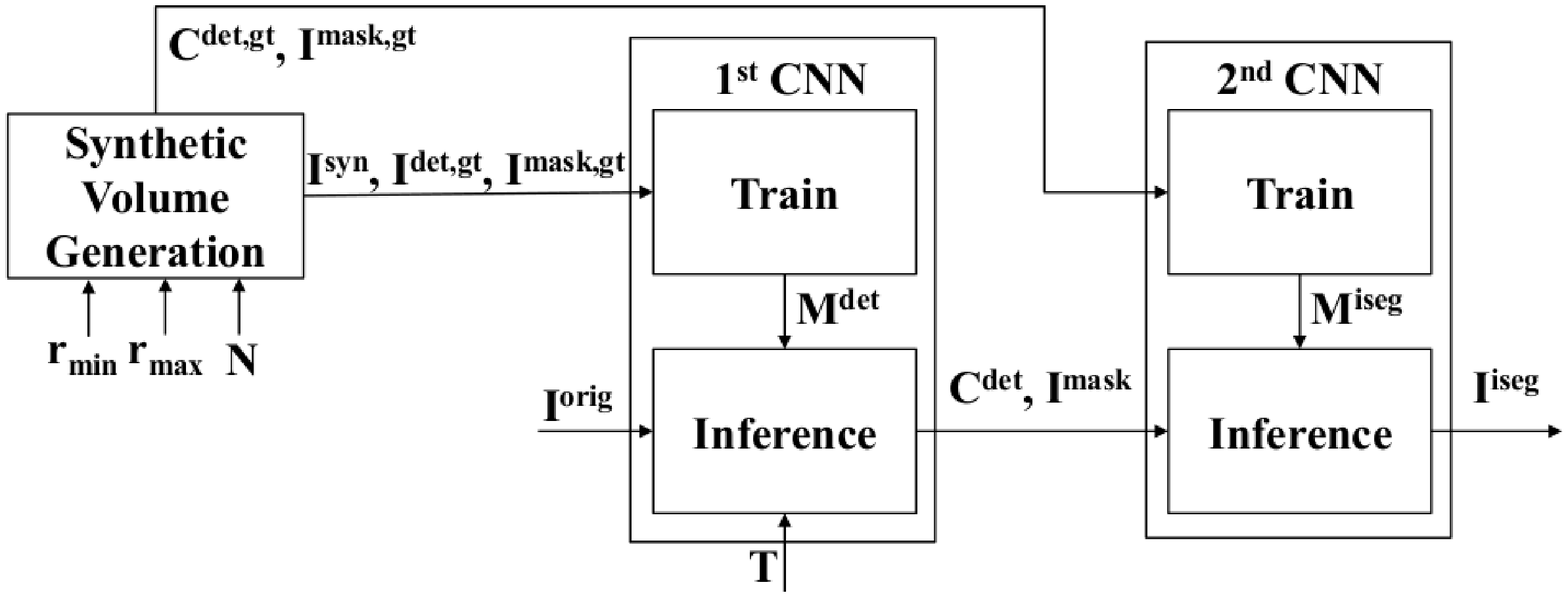,width=7cm}}
\vspace{-0.15in}
\caption{Block diagram of the proposed method}
\vspace{-0.15in}
\label{fig:block}
\end{figure}
Figure \ref{fig:block} is a block diagram of our proposed method for 3D nuclei instance segmentation.
%We use the notation we defined in \cite{ho2017b,ho2018,fu2018} to describe 3D volumes.
A 3D image volume of size $X \times Y \times Z$ is denoted as $I$ and the $p^\text{th}$ 2D focal plane image of size $X \times Y$ along the $z$-direction is denoted as $I_{z_p}$, where $p \in \{1, \dots, Z\}$.
%For example, $I_{z_{97}}^{orig}$ is the 97$^\text{th}$ focal plane image of an original volume, $I^{orig}$.
A subvolume of $I$, whose $x$-coordinate is $q_i \leq x \leq q_f$, $y$-coordinate is $r_i \leq y \leq r_f$, $z$-coordinate is $p_i \leq z \leq p_f$ is denoted as $I_{\left(q_i:q_f,r_i:r_f,p_i:p_f\right)}$, where $q_i \in \{1, \dots, X\}$, $q_f \in \{1, \dots, X\}$, $r_i \in \{1, \dots, Y\}$, $r_f \in \{1, \dots, Y\}$, $p_i \in \{1, \dots, Z\}$, and $p_f \in \{1, \dots, Z\}$.
It is required that $q_i \leq q_f$, $r_i \leq r_f$, and $p_i \leq p_f$.
%For example, $I^{iseg}_{\left(193:320,193:320,31:94\right)}$ is a subvolume of a segmented volume, $I^{iseg}$, whose $x$-coordinate is $193 \leq x \leq 320$, $y$-coordinate is $193 \leq y \leq 320$ and $z$-coordinate is $31 \leq z \leq 94$.
Lastly, a voxel is denoted as $\textbf{v}$.

Our method consists of two CNNs as shown in Figure \ref{fig:block}.
The first CNN, $M^{det}$, is used for nuclei detection and binary segmentation and the second CNN, $M^{iseg}$, is used for nuclei instance segmentation.
To segment each nucleus using the second CNN, the first CNN produces a set of coordinates of the nuclei center locations, denoted as $C^{det}$, and a nuclei mask volume denoted as $I^{mask}$. Specifically, 
$C^{det}$ consists of the centroid coordinates of components in a detection volume, $I^{det}$.
To accurately select the elements of $C^{det}$, especially when multiple nuclei are touching, the components in $I^{det}$ are chosen to have no touching regions for distinct nuclei.
%Therefore, $I^{det}$ contains components around the nuclei center locations.
The second CNN segments an individual nucleus in a 3D patch from $I^{mask}$ centered at $C^{det}$ and is color-coded to produce the final segmentation volume, $I^{iseg}$.
Note that color-coding is done to visually label each nucleus in $I^{iseg}$.
%The second CNN segments an individual nucleus in a 3D patch from $I^{mask}$ centered at $C^{det}$ and is color-coded.
%After color-coding individual nucleus, the final segmentation volume, $I^{iseg}$, is produced.
To train the two CNNs a SpCycleGAN described in \cite{fu2018} is used to generate synthetic microscopy volumes, $I^{syn}$.
Our implementation is done using PyTorch \cite{paszke2017}.

%The first CNN is used for nuclei detection and binary segmentation and has two outputs, a set of coordinates of the nuclei center location denoted as $C^{det}$ and a nuclei mask volume denoted as $I^{mask}$.
%%These two outputs are generated from an original microscopy volume, $I^{orig}$.
%$C^{det}$ is selected by finding centroid coordinates of components in a detection volume, $I^{det}$.
%To accurately select $C^{det}$ at the nuclei center locations, especially when multiple nuclei are touching, the components in $I^{det}$ are designed to have no touching regions for distinct nuclei.
%Therefore, $I^{det}$ contains components around the nuclei center locations.
%
%The second CNN is used for nuclei instance segmentation where an individual nucleus is segmented in a 3D patch from $I^{mask}$ centered at $C^{det}$ and is color-coded.
%After color-coding individual nucleus, the final segmentation volume, $I^{iseg}$, is produced.
%To train the two CNNs a SpCycleGAN described in \cite{fu2018} is used to generate synthetic microscopy volumes, $I^{syn}$.
%Our implementation is done using PyTorch \cite{paszke2017}.

\vspace{-0.05in}
\subsection{Synthetic Volume Generation}
\vspace{-0.05in}
As  indicated above, creating labeled ground truth 3D volumes is tedious.
We use the SpCycleGAN we described in \cite{fu2018} to produce synthetic microscopy 3D volumes that we use for training.
Note we do not need any actual ground truth volumes to use the approach described in this section.
Synthetic microscopy volumes, $I^{syn}$, nuclei mask ground truth volumes, $I^{mask,gt}$, and detection ground truth volumes, $I^{det,gt}$, need to be generated.
We start by generating a random 3D nuclei mask volume and then use it to generate the synthetic volume.
To generate $I^{mask,gt}$ we develop two approaches: the first approach produces $N$ synthetic spherical nuclei and the second approach produces elliptical nuclei based on nuclei structures in $I^{orig}$.
%In our approach synthetic nuclei can be generated  with either spherical or  elliptical based nuclei  shapes in $I^{orig}$ with $N$ synthetic nuclei.
For the first approach the $i^\text{th}$ synthetic nuclei, $I^{mask,i}$, is generated as a sphere with a randomly selected radius, $r_i$, between $r_{min}$ and $r_{max}$, and centered at a randomly selected coordinate, $C^{det,i}$, where $1 \leq i \leq N$.
%\vspace{-0.05in}
%\begin{equation}
	%I^{mask,i}(\textbf{v}) = 
%\begin{cases}
	%1,& \text{if} \quad ||\textbf{v}-C^{det,i}||_2^2 < r_i^2\\
%0,& \text{otherwise}
%\end{cases}
%\vspace{-0.05in}
%\label{eq:maski}
%\end{equation}
Simultaneously, the $i^\text{th}$ central region, $I^{det,i}$, is generated where a central region of a nucleus is defined as a sphere inside the nucleus where the centroid of the central region matches to the centroid of the nucleus.
%Simultaneously, the $i^\text{th}$ central region, $I^{det,i}$, is generated.
%We define a central region of a nucleus as a sub-region inside the nucleus where the centroid of the sub-region matches to the centroid of the nucleus.
%\vspace{-0.05in}
%\begin{equation}
	%I^{det,i}(\textbf{v}) = 
%\begin{cases}
	%1,& \text{if} \quad ||\textbf{v}-C^{det,i}||_2^2 < \left(\frac{r_i}{2}\right)^2\\
%0,& \text{otherwise}
%\end{cases}
%\vspace{-0.05in}
%\label{eq:deti}
%\end{equation}
We intentionally set the radius of $I^{det,i}$ to be $\frac{r_i}{2}$ to avoid multiple connected central regions although their corresponding synthetic nuclei may be touching.
Once $N$ synthetic nuclei and their central regions are produced, they are 
added to $I^{mask,gt}$ and $I^{det,gt}$ sequentially where $I^{mask,gt}$ and $I^{det,gt}$ are 
initialized to zero.
If $I^{mask,i}$ overlaps with any previous synthetic nuclei in $I^{mask,gt}$, then $I^{mask,i}$ and $I^{det,i}$ are not added to $I^{mask,gt}$ and $I^{det,gt}$, respectively.

For the second approach $I^{mask,i}$ is generated as an ellipsoid with randomly and independently selected three semi-axes between $r_{min}$ and $r_{max}$, randomly rotated in $x$, $y$, and $z$-axes, and centered at a randomly selected coordinate, $C^{det,i}$.
In our experiments we used both approaches for generating synthetic images.

\begin{figure}[htb]
\centering
\subfigure[]
{
	\epsfig{figure=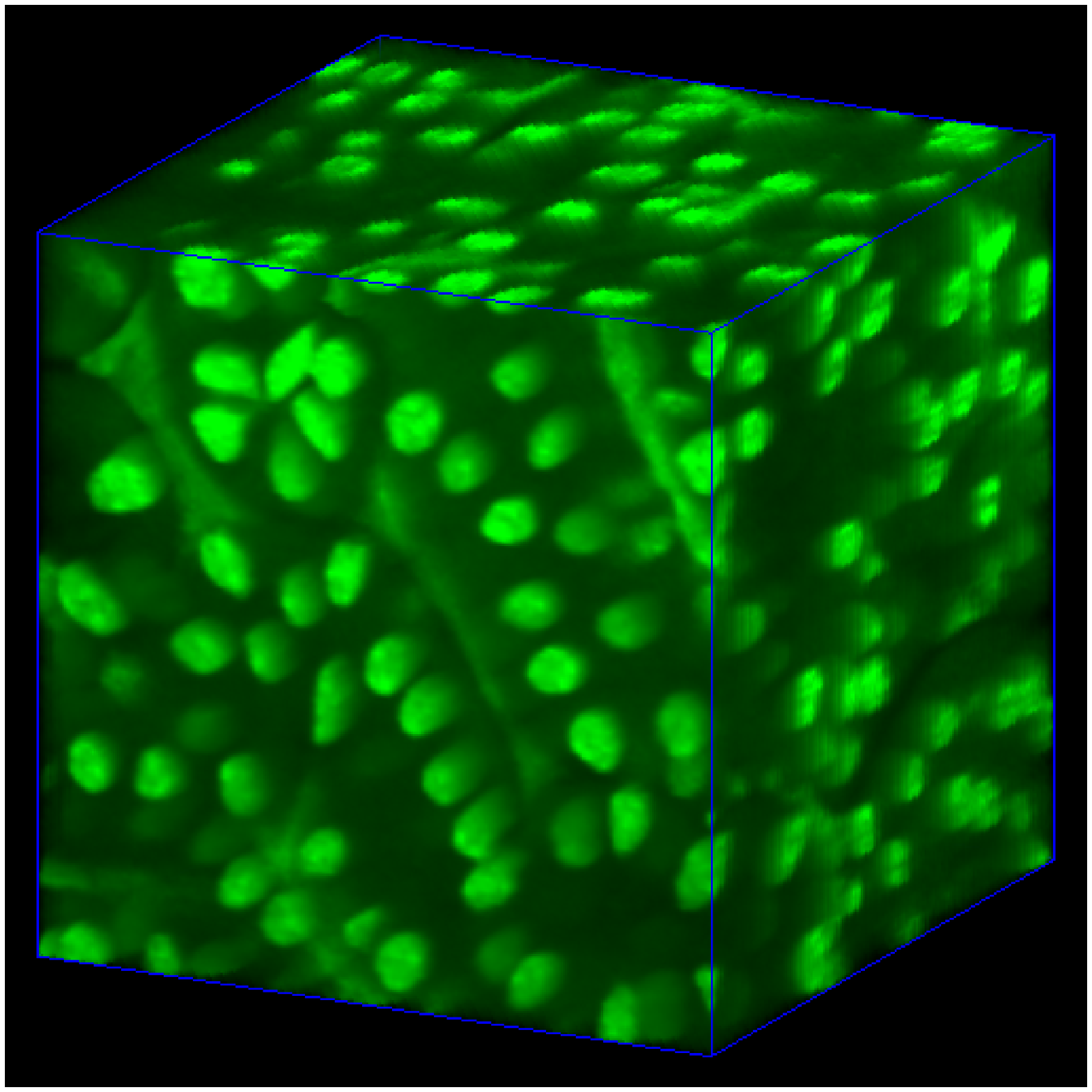,width=1.9cm}
}
\subfigure[]
{
	\epsfig{figure=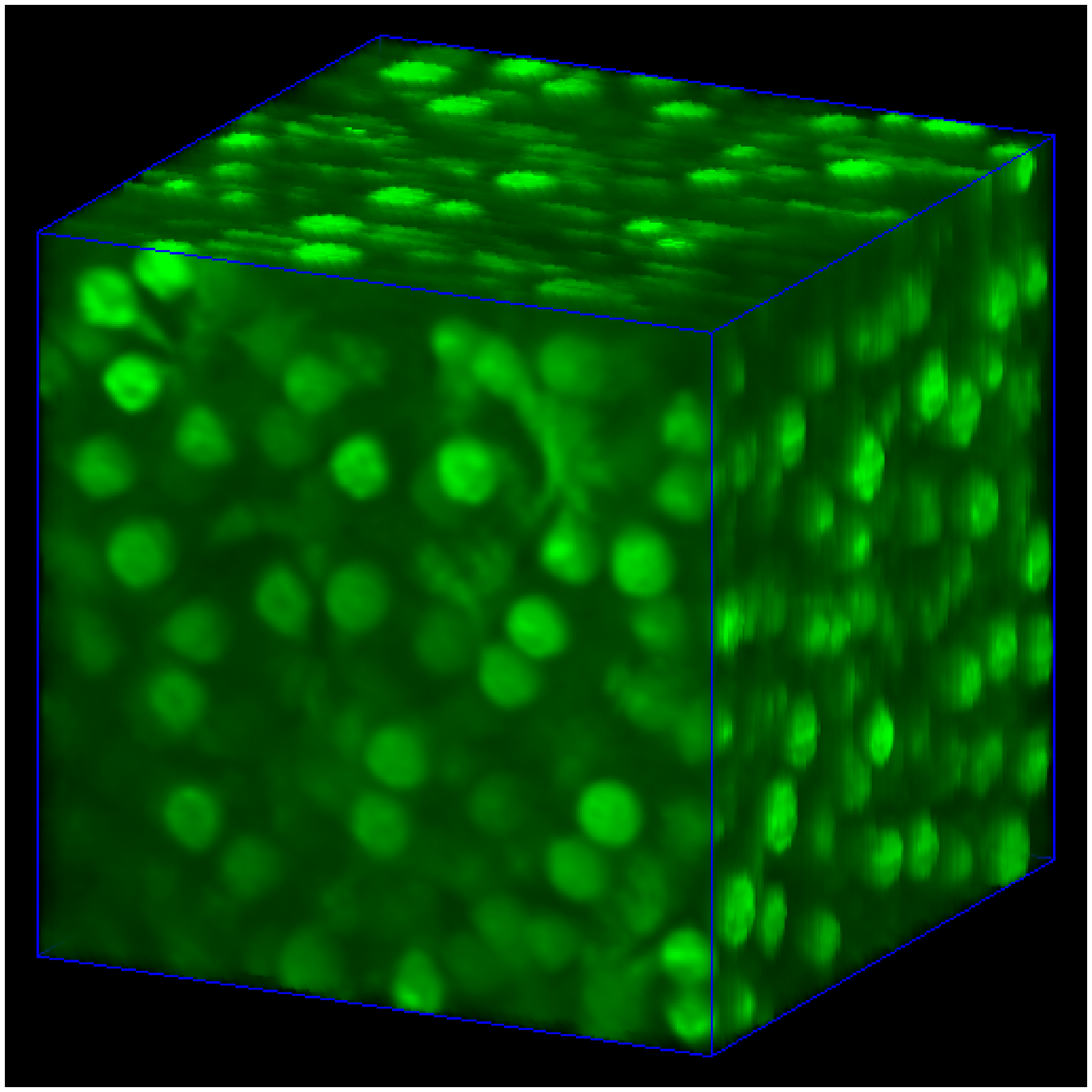,width=1.9cm}
}
\subfigure[]
{
	\epsfig{figure=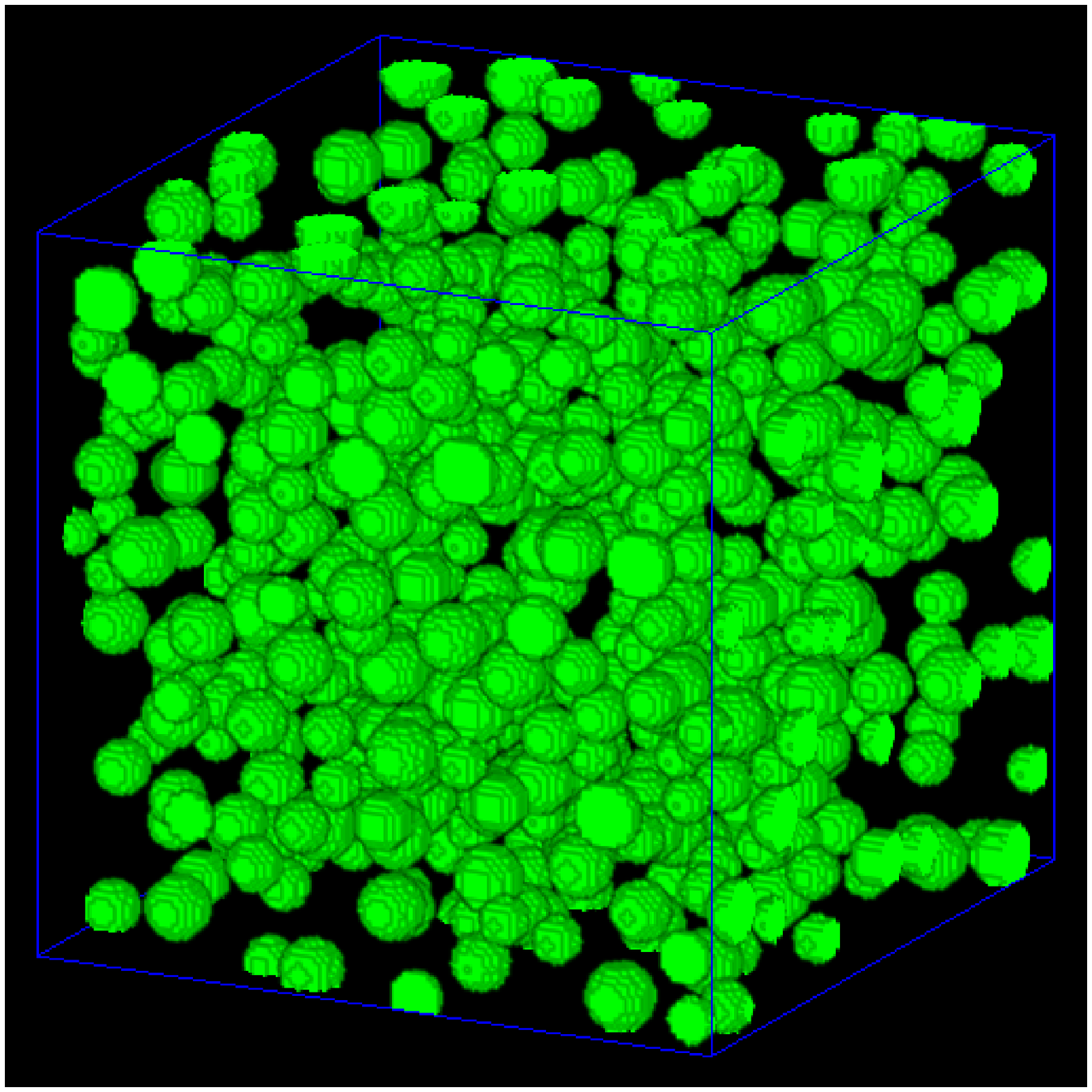,width=1.9cm}
}
\subfigure[]
{
	\epsfig{figure=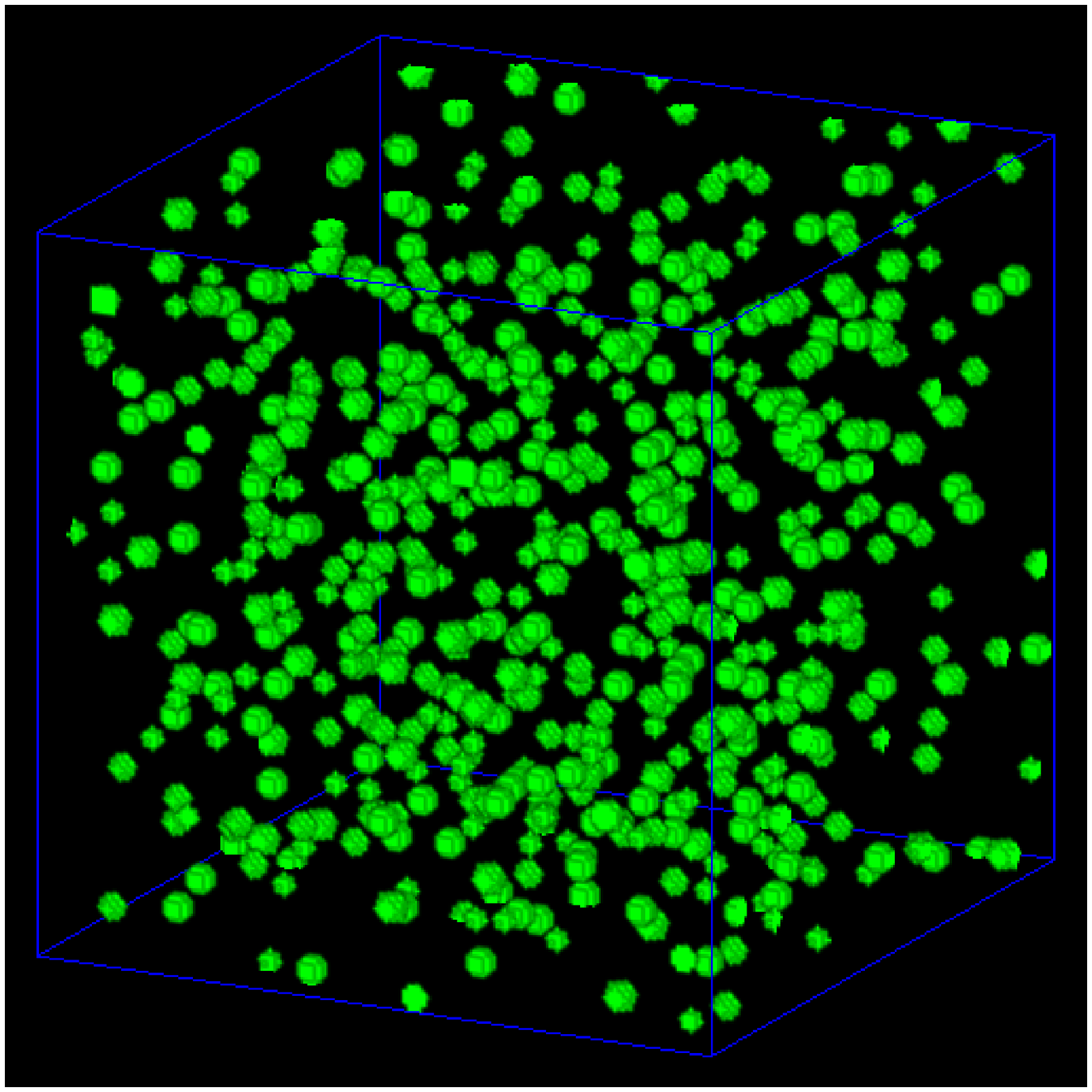,width=1.9cm}
}
\vspace{-0.15in}
\caption{Example volumes (a) real microscopy volume (b) synthetic microscopy volume (c) nuclei mask ground truth volume (d) detection ground truth volume}
%\caption{Examples of a real microscopy volume, a synthetic microscopy volume and synthetic ground truth volumes (a) real microscopy volume (b) synthetic microscopy volume (c) nuclei mask ground truth volume (d) detection ground truth volume}
\vspace{-0.15in}
\label{fig:synthetic}
\end{figure}

Once the nuclei mask ground truth volume, $I^{mask,gt}$, and the detection ground truth volume, $I^{det,gt}$, are generated, we use the SpCycleGAN to generate the corresponding synthetic volume, $I^{syn}$.
For our experiments we generated 20 sets of synthetic volumes with a size of $128 \times 128 \times 128$.
Figure \ref{fig:synthetic} shows examples of a real microscopy volume, a synthetic microscopy volume, and synthetic ground truth volumes visualized by Voxx \cite{clendenon2002}, respectively.
\vspace{-0.15in}

\vspace{-0.05in}
\subsection{Nuclei Detection and Binary Segmentation}
\vspace{-0.15in}
\begin{figure}[htb]
\centerline{\epsfig{figure=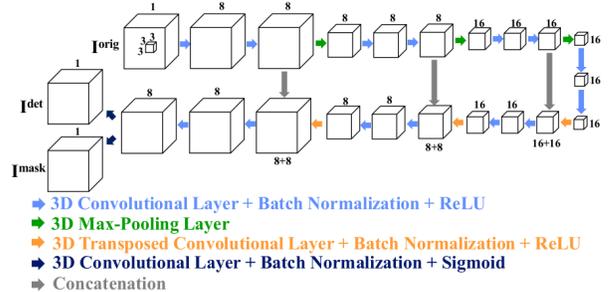,width=8cm}}
\vspace{-0.15in}
\caption{Our first CNN architecture (see Figure \ref{fig:block})}
\vspace{-0.15in}
\label{fig:CNN1}
\end{figure}

Our first CNN used for nuclei detection and binary segmentation outputs nuclei center locations, $C^{det}$, and a nuclei mask volume, $I^{mask}$ (Figure \ref{fig:block}).
This CNN is shown in more detail in Figure \ref{fig:CNN1} and
uses a modified 3D U-Net architecture \cite{cicek2016}.
%has an encoder-decoder architecture which is modified from 3D U-Net \cite{cicek2016}.
$C^{det}$ can be selected by finding centroids of elements of $I^{det}$.
To avoid false-detection, labels in $I^{det}$ are labeled as background if $I^{mask}$ at the same voxel locations are labeled as background.
Also, components with the number of voxels less than $T$ are not considered in order to remove noise.
A 3D convolutional layer consists of a convolutional operation with a $3 \times 3 \times 3$ kernel with 1 voxel padding, 3D batch normalization, and a rectified-linear unit (ReLU) activation function.
Note that the Sigmoid function is used as an activation function for the last convolutional layers.
In the encoder, 3D max-pooling layer uses $2 \times 2 \times 2$ window with a stride of 2. 
In the decoder, a 3D transposed convolutional layer followed by 3D batch normalization and a ReLU activation function is used.
In addition, concatenation transfers feature maps from the encoder to the decoder.
The size of input/output volumes are $64 \times 64 \times 64$.
If the size of $I^{orig}$ is larger than $64 \times 64 \times 64$, then a 3D window with size of $64 \times 64 \times 64$ is moved in the $x$, $y$, and $z$-directions until the entire $I^{orig}$ is processed \cite{ho2017b}.
During  training, the Adam optimizer \cite{kingma2014} is used with a learning rate of 0.001.
The training loss function is a sum of the Binary Cross Entropy (BCE) loss of the detection volume and the BCE loss of the nuclei mask volume.
The BCE loss, $L^{BCE}$, is defined as
$L^{BCE}(I^{out},I^{gt}) = -\frac{1}{V}\sum_{\textbf{v}=1}^{V} \big(I^{gt}(\textbf{v})\log I^{out}(\textbf{v}) + (1-I^{gt}(\textbf{v}))\log (1-I^{out}(\textbf{v}))\big)$
%\vspace{-0.05in}
%\begin{align}
%	L^{BCE}(I^{out},I^{gt}) = &-\frac{1}{V}\sum_{\textbf{v}=1}^{V} \big(I^{gt}(\textbf{v})\log I^{out}(\textbf{v}) \nonumber \\
%	& + (1-I^{gt}(\textbf{v}))\log (1-I^{out}(\textbf{v}))\big)
%\vspace{-0.05in}
%\label{eq:BCE}
%\end{align}
where $I^{out}$ is the output volume, $I^{gt}$ is the ground truth volume, and $V$ is the total number of voxels in the volume.
For the training set, we used 160 synthetic volumes with a size of $64 \times 64 \times 64$. 
Each synthetic volume with a size of $128 \times 128 \times 128$ generated in the synthetic volume generation stage is divided into 8 volumes with a size of $64 \times 64 \times 64$.

\vspace{-0.05in}
\subsection{Nuclei Instance Segmentation}
\vspace{-0.15in}
\begin{figure}[htb]
\centerline{\epsfig{figure=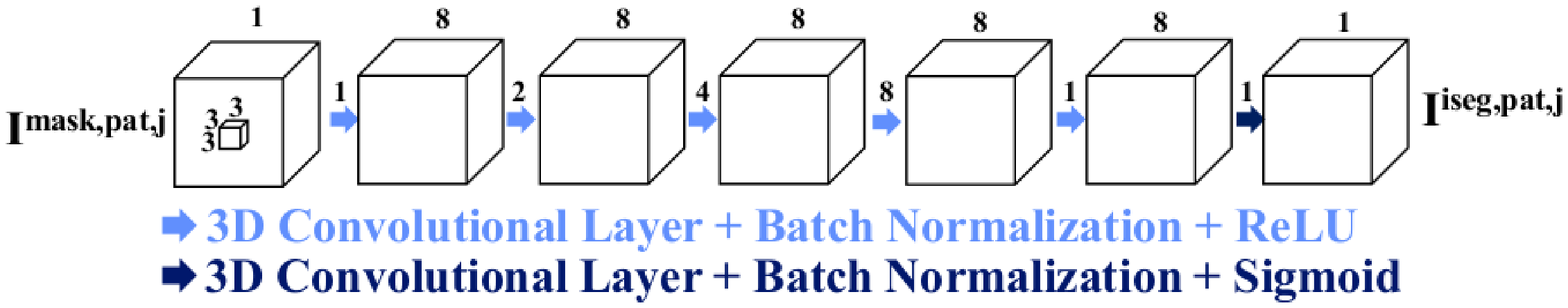,width=8cm}}
\vspace{-0.15in}
\caption{Our second CNN architecture (see Figure \ref{fig:block})}
\vspace{-0.15in}
\label{fig:CNN2}
\end{figure}

The goal of our method is nuclei instance segmentation which is segmenting individual detected nuclei with distinct labels.
Therefore, the last step is to segment each nucleus in $I^{mask}$ at a detected coordinate, $C^{det}$, using our second CNN shown in Figure \ref{fig:CNN2}.
First, the $j^\text{th}$ nucleus is cropped and included in a 3D patch with a size of $32 \times 32 \times 32$ from $I^{mask}$ centered at $C^{det,j}$, denoted as $I^{mask,pat,j}$.
Then the second CNN segments only the $j^\text{th}$ nucleus in $I^{mask,pat,j}$ and removes other nuclei structures partially included in the patch.
Here, we denote the segmented nucleus as $I^{iseg,pat,j}$.
Once the $j^\text{th}$ nucleus is segmented, it is color-coded and inserted in $I^{iseg}$ where the center location of $I^{iseg,pat,j}$ lies at $C^{det,j}$.

The second CNN in Figure \ref{fig:block}, $M^{iseg}$, consists of a series of 3D convolutional layers.
We use dilated convolutions \cite{yu2016} to have receptive field larger than the size of the patch.
From the $k^\text{th}$ feature map, $I^k$, with a convolution filter, $h$, the $(k+1)^\text{th}$ feature map, $I^{k+1}$, is generated using a $d$-dilated convolution at a voxel $\textbf{v}$ as
$I^{k+1}(\textbf{v}) = \sum_{\textbf{u}}I^k(\textbf{v}-d\textbf{u})h(\textbf{u})$ 
%\vspace{-0.05in}
%\begin{equation}
%	I^{k+1}(\textbf{v}) = \sum_{\textbf{u}}I^k(\textbf{v}-d\textbf{u})h(\textbf{u})
%\label{eq:dilation}
%\vspace{-0.05in}
%\end{equation}
where $d$ is known as the dilation factor.
%Dilated convolutions are used to increase the receptive fields.
Figure \ref{fig:CNN2} shows the dilation factors for the convolutional layers such that the final receptive field is larger than $32 \times 32 \times 32$.
Note the kernel size for the last convolutional layer of the second CNN is $1 \times 1 \times 1$.
%Table \ref{tab:dilation} shows the dilation factors for the convolutional layers such that the final receptive field is larger than $32 \times 32 \times 32$.
During training, the Adam optimizer \cite{kingma2014} is used with a learning rate of 0.001.
The BCE loss is used as the training loss function.
300 patches from $I^{mask,gt}$ centered at $C^{det,gt}$ are used for the training.

%\vspace{-0.2in}
%\begin{table}[htb]
%\centering
%{
%\caption{Dilation factors for the convolutional layers and their corresponding receptive fields}
%\begin{tabular}{| c | c | c | c |}
	%\hline
	%Conv. Layer & Filter Size & Dilation Factor & Receptive Field\\
	%\hline
  %1$^\text{st}$ layer & $3 \times 3 \times 3$ & 1 & $3 \times 3 \times 3$ \\
	%\hline
	%2$^\text{nd}$ layer & $3 \times 3 \times 3$ & 2 & $7 \times 7 \times 7$ \\
	%\hline
	%3$^\text{rd}$ layer & $3 \times 3 \times 3$ & 4 & $15 \times 15 \times 15$ \\
	%\hline
	%4$^\text{th}$ layer & $3 \times 3 \times 3$ & 8 & $31 \times 31 \times 31$ \\
	%\hline
	%5$^\text{th}$ layer & $3 \times 3 \times 3$ & 1 & $33 \times 33 \times 33$ \\
	%\hline
	%6$^\text{th}$ layer & $1 \times 1 \times 1$ & 1 & $33 \times 33 \times 33$ \\
	%\hline
%\end{tabular}
%%\vspace{-0.2in}
%\label{tab:dilation}
%}
%\end{table}

\vspace{-0.05in}
\section{Experimental Results}
\begin{figure}[htb]
\centering
\subfigure[]
{
	\epsfig{figure=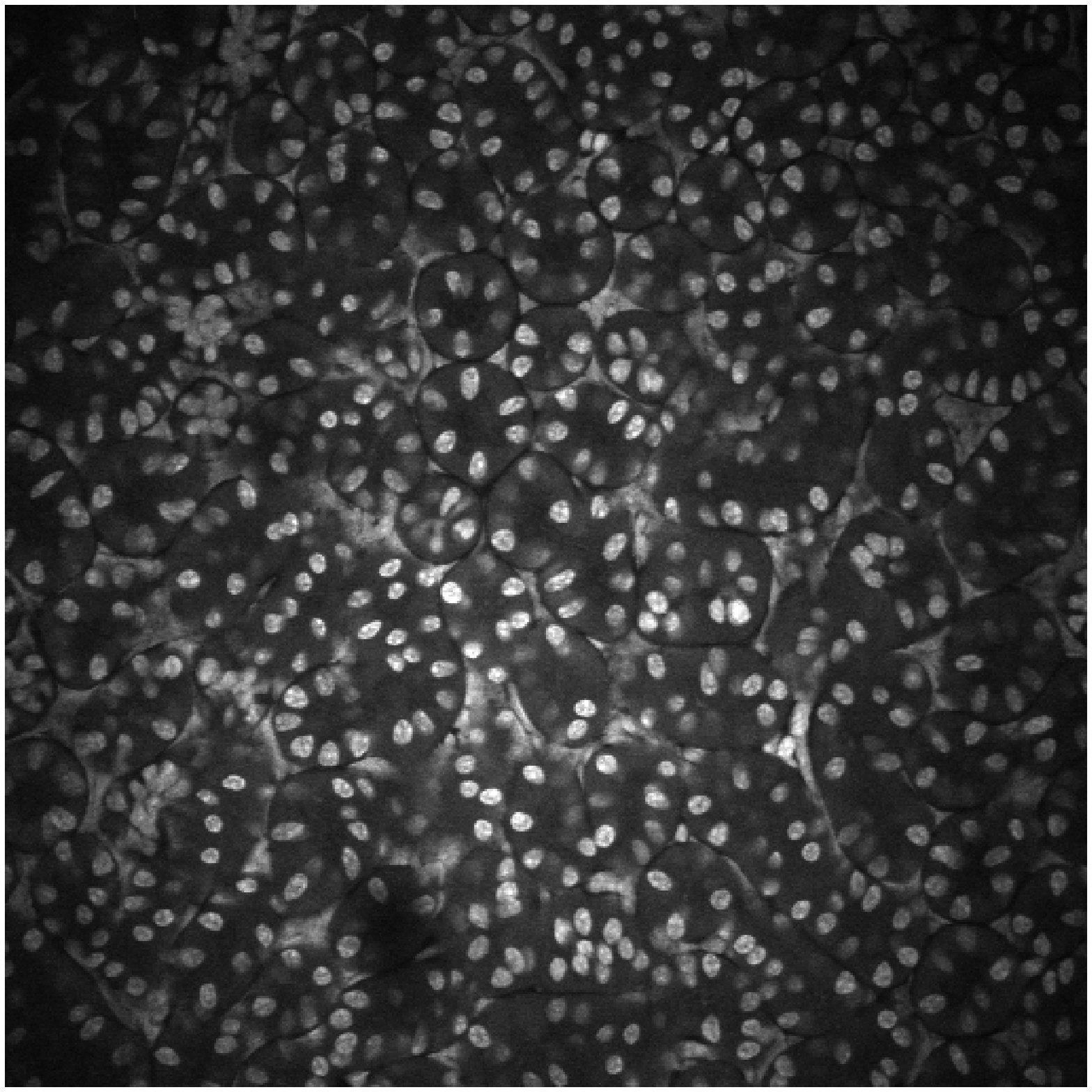,width=2.6cm}
}
\subfigure[]
{
	\epsfig{figure=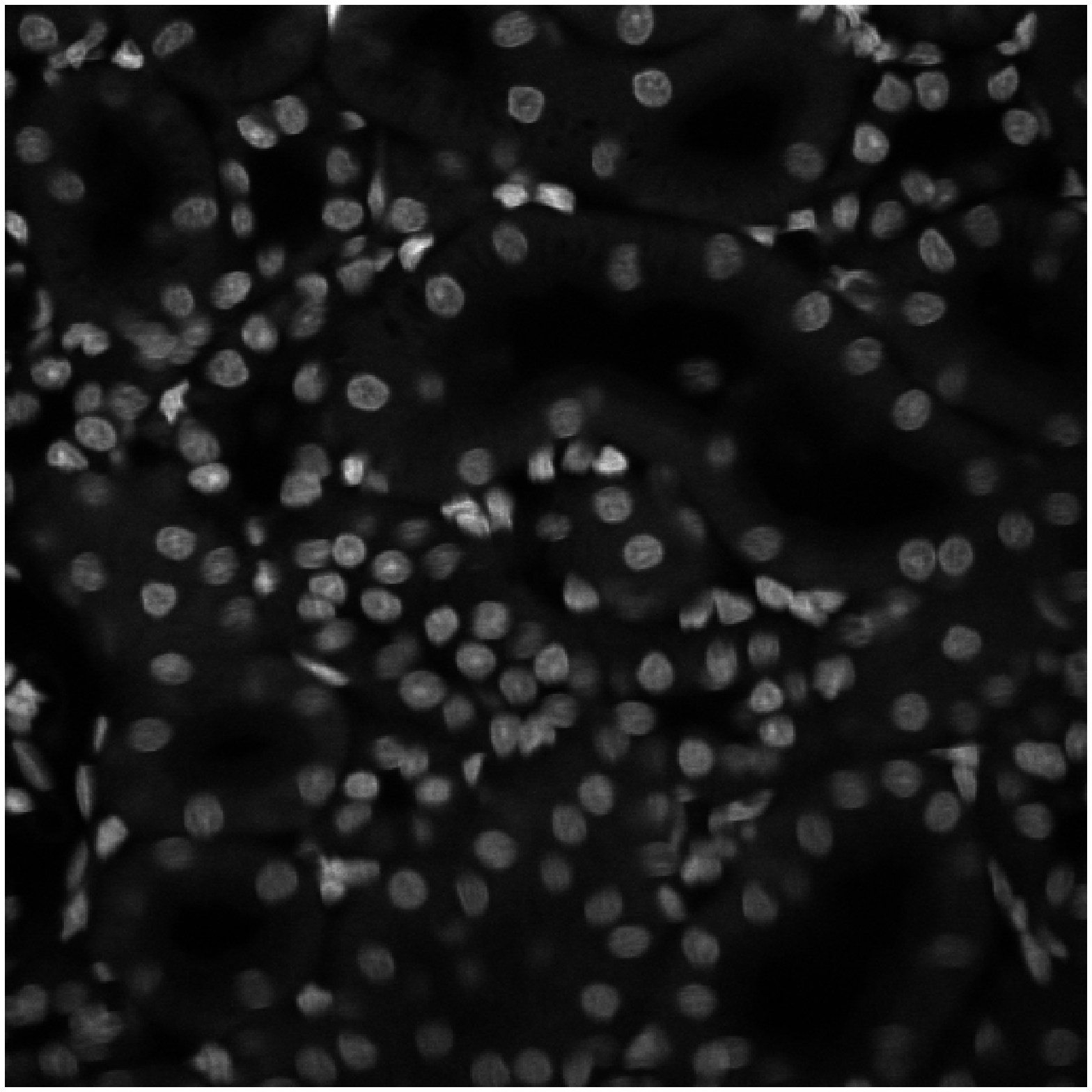,width=2.6cm}
}
\subfigure[]
{
	\epsfig{figure=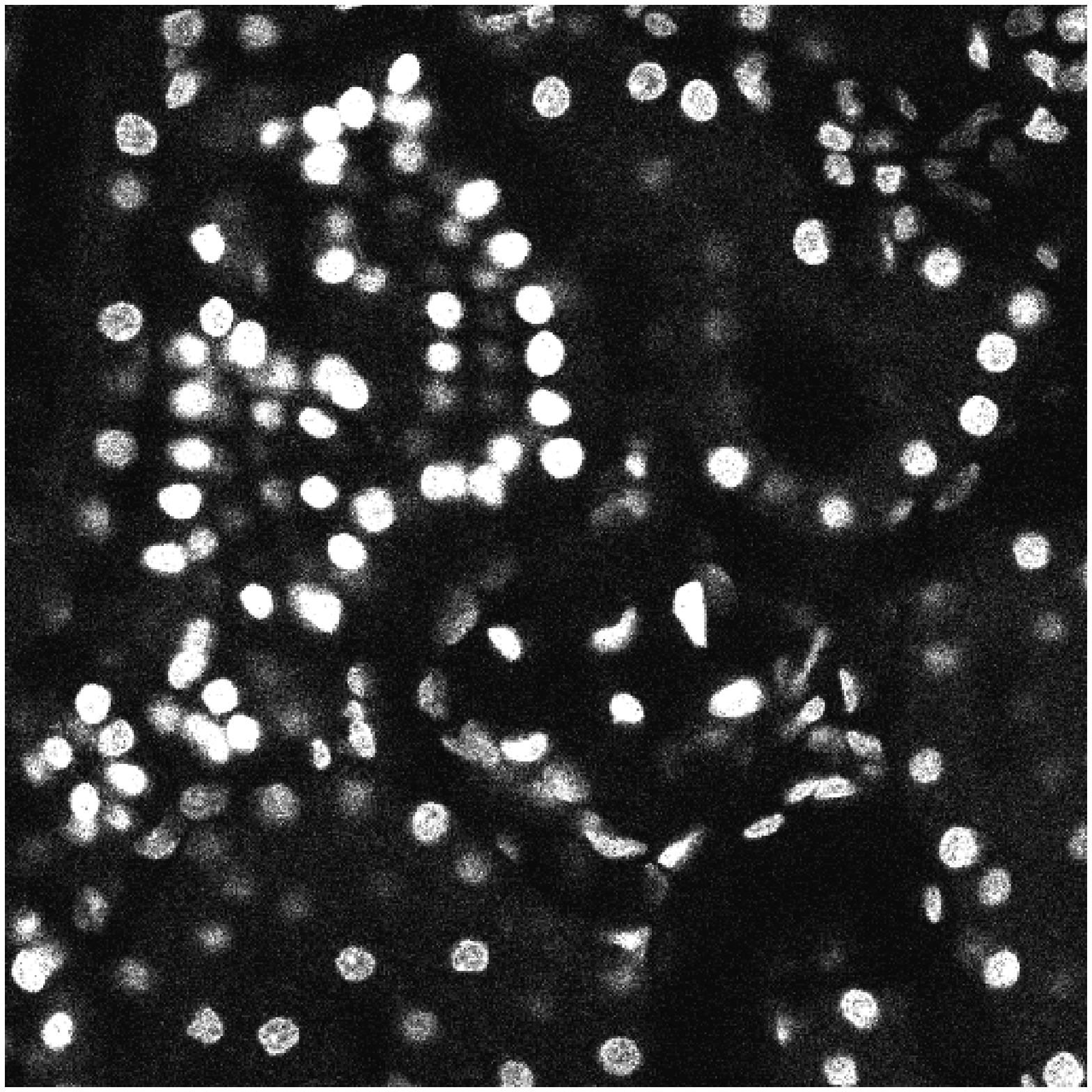,width=2.6cm}
}
\subfigure[]
{
	\epsfig{figure=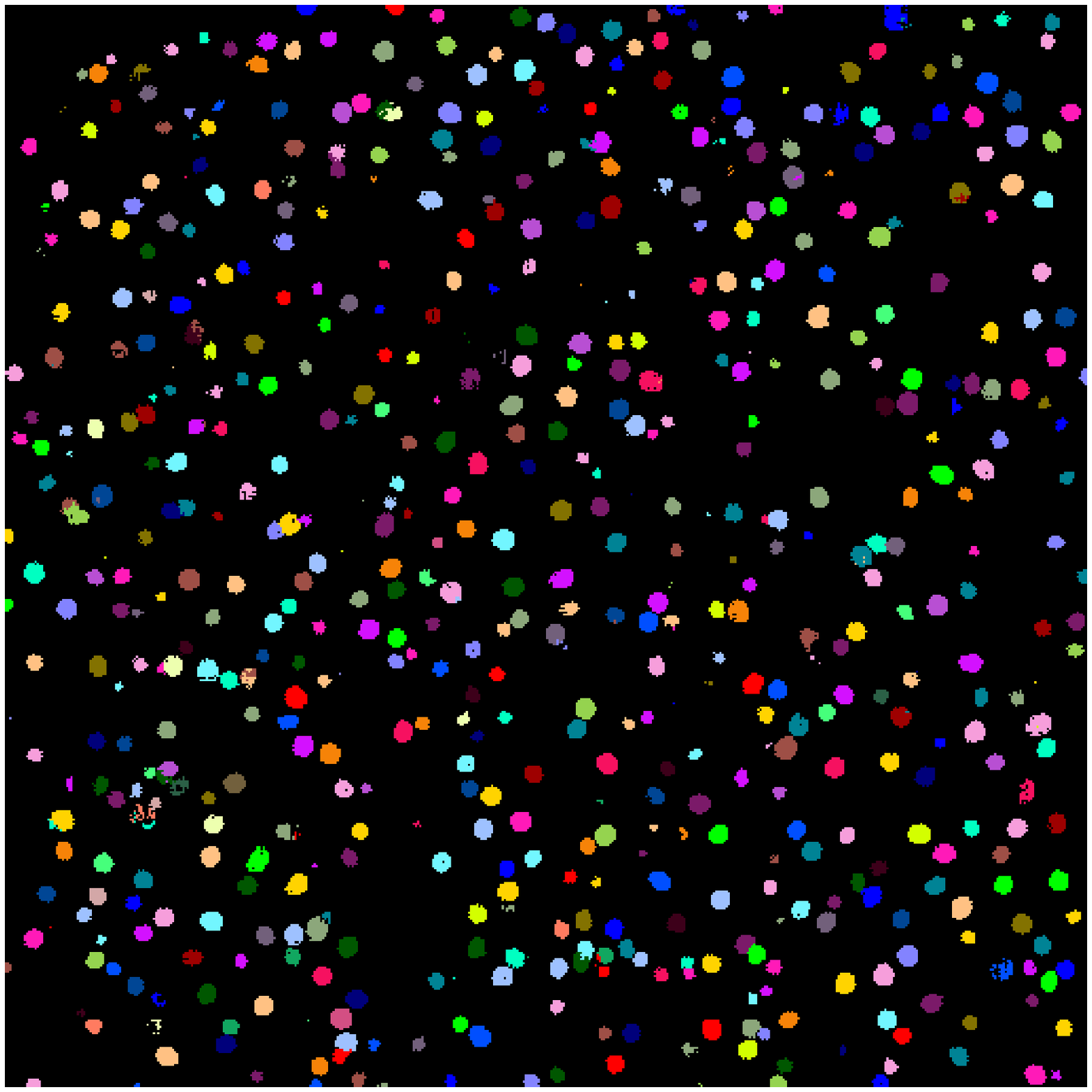,width=2.6cm}
}
\subfigure[]
{
	\epsfig{figure=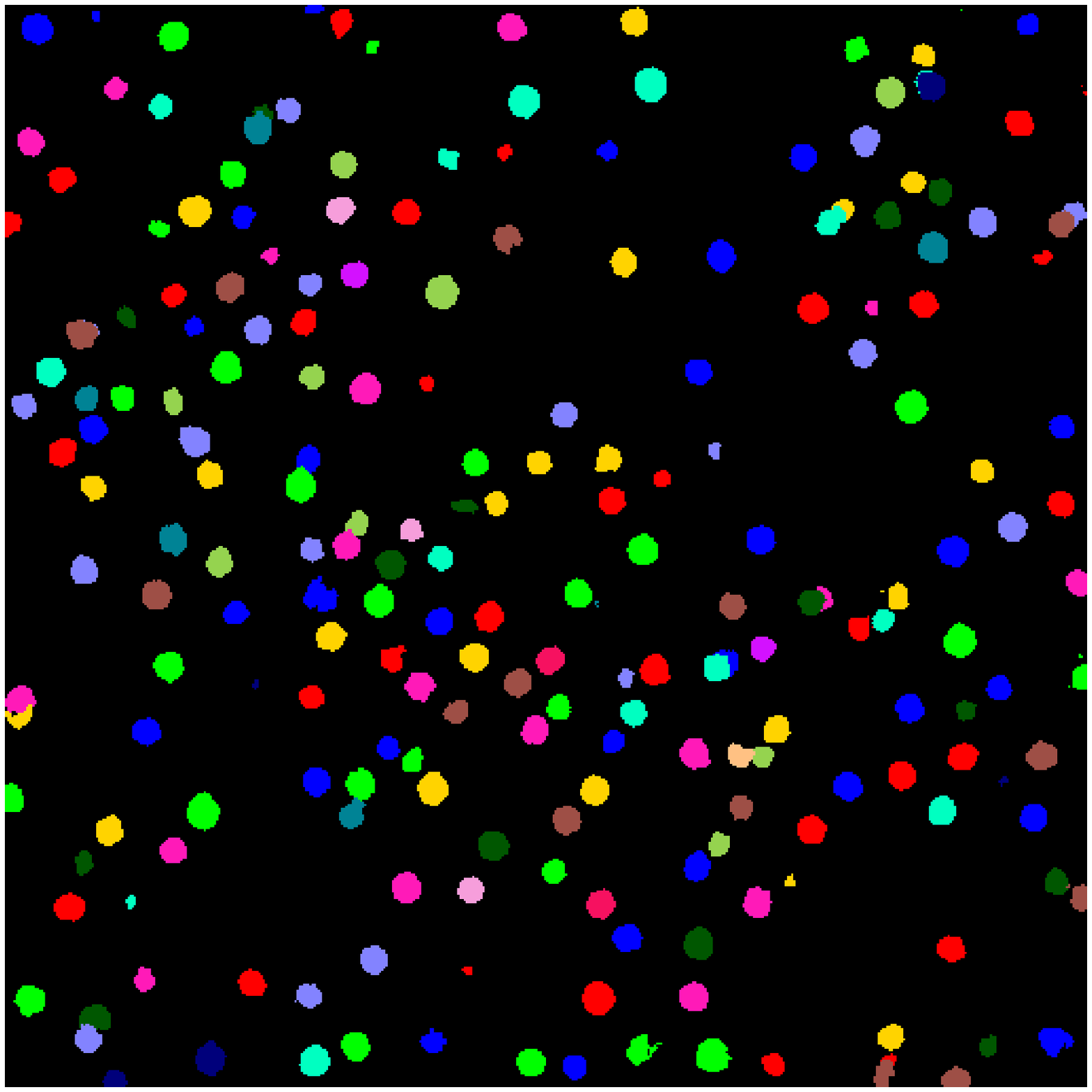,width=2.6cm}
}
\subfigure[]
{
	\epsfig{figure=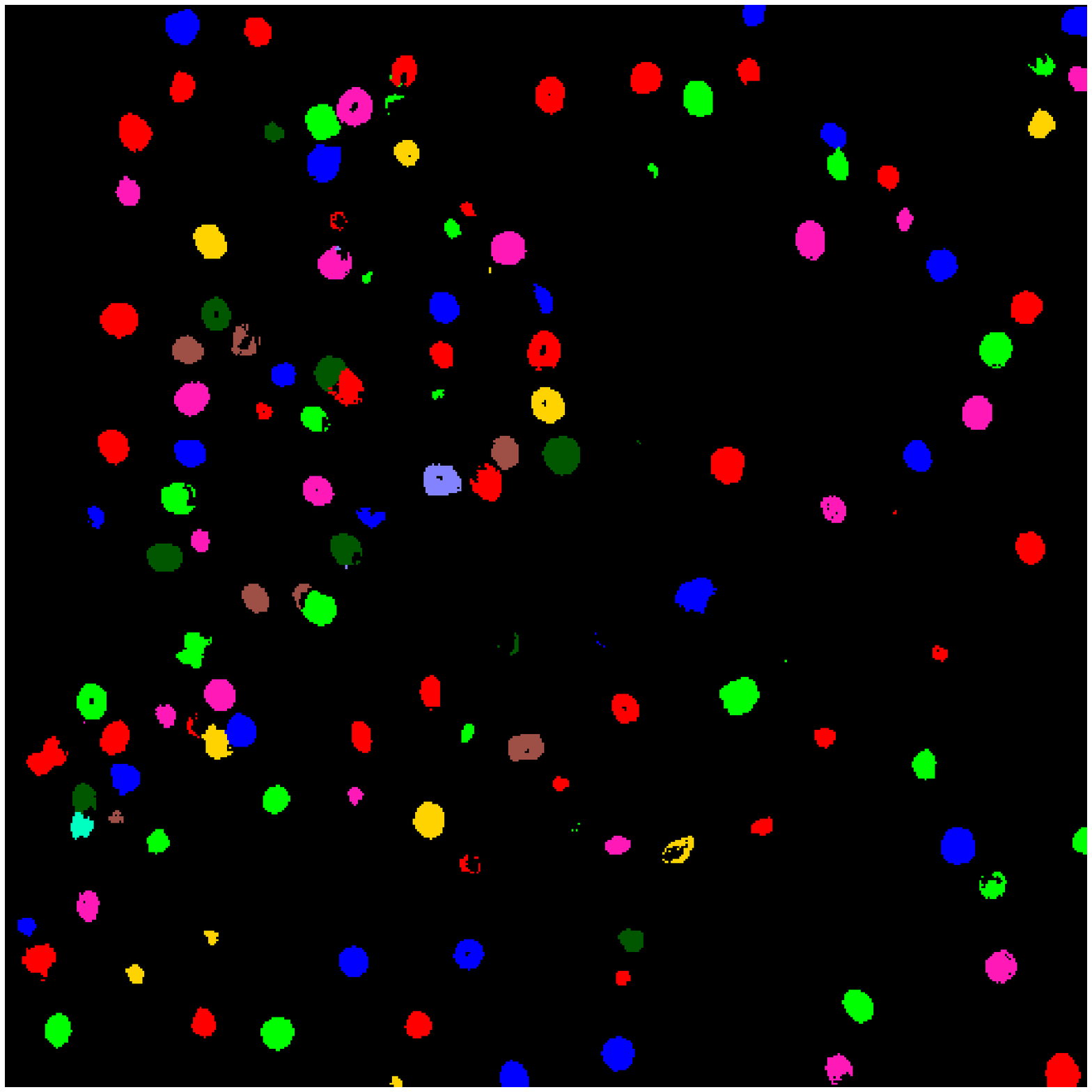,width=2.6cm}
}
\vspace{-0.15in}
\caption{Original and segmented images in Data-I, Data-II, and Data-III (a) $I^{orig}_{z_{97}}$ for Data-I, (b) $I^{orig}_{z_{403}}$ for Data-II, (c) $I^{orig}_{z_{14}}$ for Data-III, (d) $I^{iseg}_{z_{97}}$ for Data-I, (e) $I^{iseg}_{z_{403}}$ for Data-II, (f) $I^{iseg}_{z_{14}}$ for Data-III}
\vspace{-0.15in}
\label{fig:results}
\end{figure}

\vspace{-0.05in}
\label{sec:result}
Our method is tested on three rat kidney data sets.
All data sets consist of gray scale images of size $X = 512 \times Y = 512$.
Data-I consists of $Z = 512$ images, Data-II of $Z = 415$, and Data-III of $Z = 45$. 
% data-I: water-scale-mount, data-II: 3color, data-III: cliff-kidney
To match resolution in $z$-direction to resolution in $x$ and $y$-directions, Data-II is downsampled in $z$-direction by a factor of 2 and Data-III is linearly interpolated in $z$-direction by a factor of 2.
$r_{min} = 4$, $r_{max} = 6$, and $N = 1000$ with a spherical model and $T = 10$ for Data-I, $r_{min} = 6$, $r_{max} = 9$, and $N = 200$ with an ellipsoidal model and $T = 20$ for Data-II, and $r_{min} = 6$, $r_{max} = 9$, and $N = 50$ with a spherical model and $T = 30$ for Data-III are used, respectively.
Note that the size of synthetic nuclei for Data-I is small, so the size of patches during nuclei instance segmentation is reduced to $16 \times 16 \times 16$ and the fourth convolutional layer in $M^{iseg}$ is removed.
Figure \ref{fig:results} shows original images and segmented images for Data-I, Data-II, and Data-III.

\vspace{-0.1in}

Our method was compared to other segmentation methods using object-wise evaluation criterion \cite{powers2011}. %fawcett2006
%Our method was compared to other segmentation methods using voxel-wise and object-wise evaluation criteria \cite{powers2011}.%fawcett2006
The other segmentation methods include Squassh \cite{rizk2014}, watershed \cite{vincent1991}, our previous detection and segmentation method \cite{ho2018} that we will denote as Purdue1, and our previous segmentation method using a SpCycleGAN \cite{fu2018} that we will denote as Purdue2.
Note our method in \cite{fu2018} generates binary segmentation masks but cannot label nuclei distinctly.
To label touching nuclei distinctly, we added a post-processing step in Purdue2 using morphological operations with a 3D erosion, a 3D connected component for color-coding, and a 3D dilation with a sphere of radius of 1 used as the structuring element.
%For the voxel-wise evaluation, we define $\text{accuracy} = \frac{n_\text{TP}+n_\text{TN}}{n_\text{total}}$, $\text{Type-I error} = \frac{n_\text{FP}}{n_\text{total}}$, $\text{Type-II error} = \frac{n_\text{FN}}{n_\text{total}}$, where $n_\text{TP}$, $n_\text{TN}$, $n_\text{FP}$, $n_\text{FN}$, $n_\text{total}$ are the number of true positive voxels (voxels correctly labeled as nuclei), true negative voxels (voxels correctly labeled as background), false positive voxels (voxels wrongly labeled as nuclei), false negative voxels (voxels wrongly labeled as background), and the total number of voxels in a volume, respectively.
For the object-wise evaluation, Precision ($P$), Recall ($R$), and F1 score ($F1$) are defined as $P = \frac{N_{TP}}{N_{TP} + N_{FP}}$, $R = \frac{N_{TP}}{N_{TP} + N_{FN}}$, and $F1 = \frac{2PR}{P+R}$, where $N_{TP}$, $N_{FP}$, and $N_{FN}$ are the number of true positive objects, the number of false positive objects, and the number of false negative objects, respectively.
A segmented nucleus is defined as a true positive object if it intersects at least 50\% of the corresponding ground truth nucleus.
Otherwise, it is defined as a false positive object.
A ground truth nucleus is defined as a false negative object if it intersects less than 50\% of the corresponding segmented nucleus or there is no corresponding segmented nucleus.
In our evaluation, we generated a 3D ground truth volume, $I^{gt}_{\left(193:320,193:320,31:94\right)}$, using ITK-SNAP \cite{yushkevich2006} from Data-I with size of $128 \times 128 \times 64$ containing 283 nuclei.
Note that any components whose number of voxels is less than 50 are removed on $I^{iseg}_{\left(193:320,193:320,31:94\right)}$ and $I^{gt}_{\left(193:320,193:320,31:94\right)}$ to remove partially included nuclei on the boundary of the subvolume.

\begin{figure}[htb]
\centering
\subfigure[]
{
	\epsfig{figure=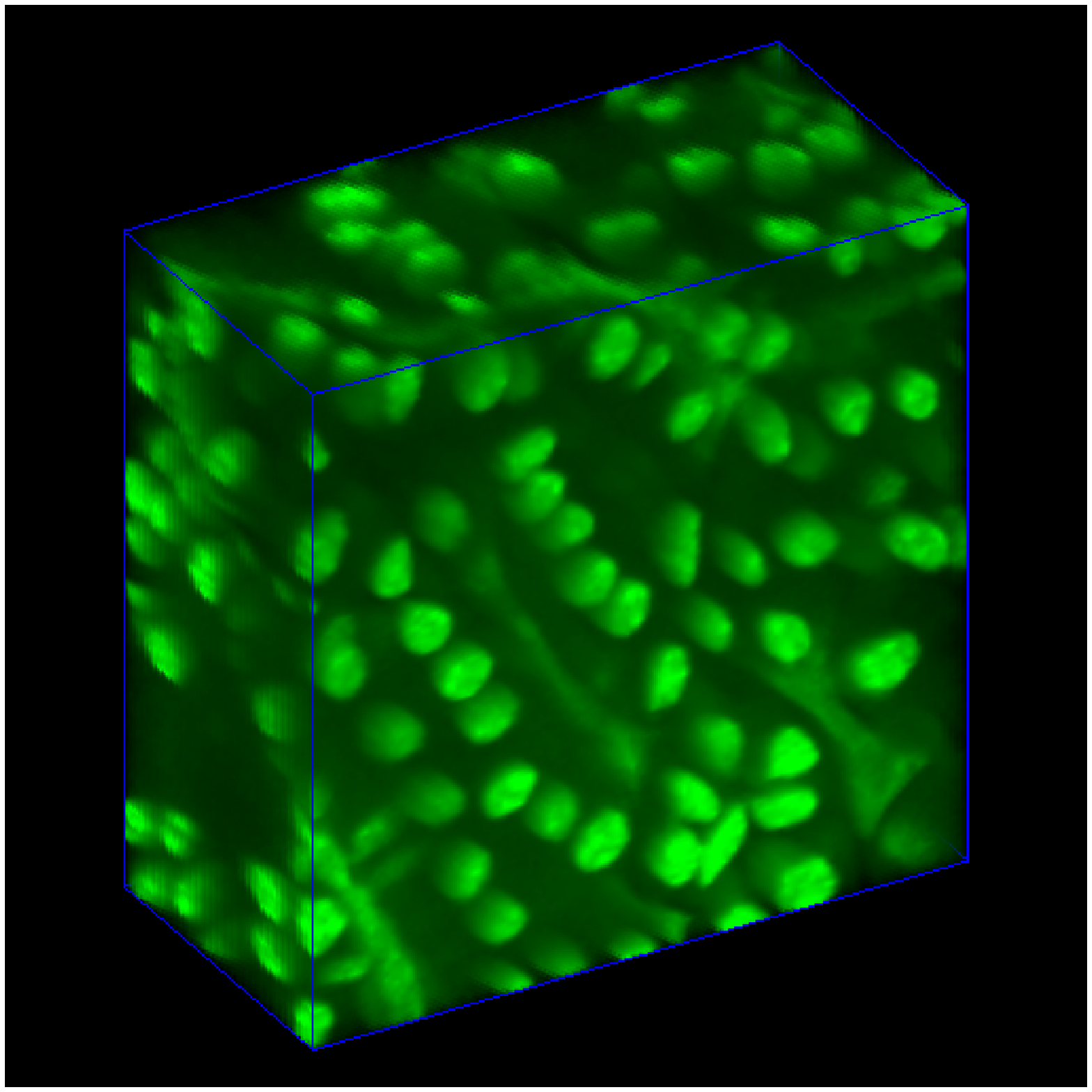,width=1.9cm}
}
\subfigure[]
{
	\epsfig{figure=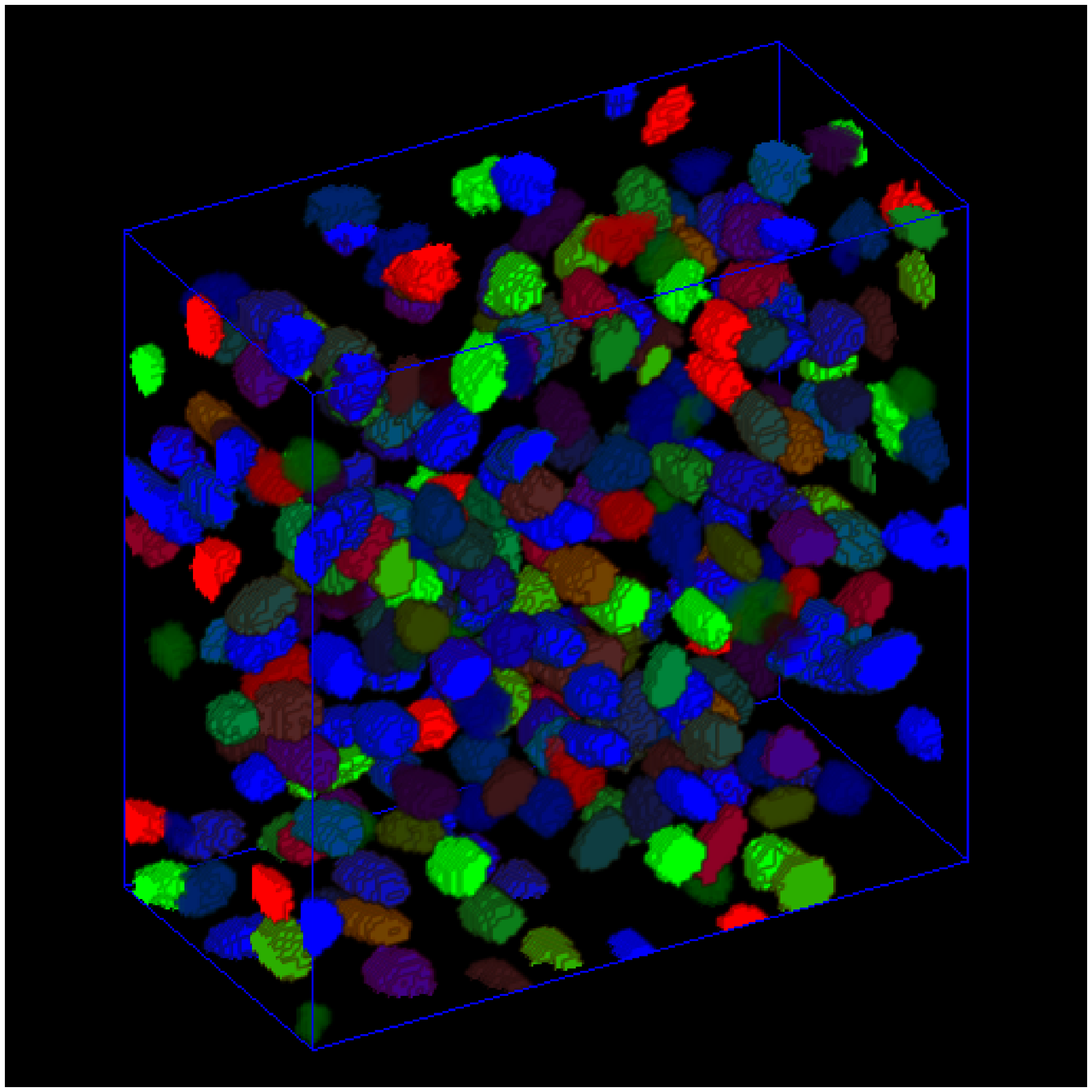,width=1.9cm}
}
\subfigure[]
{
	\epsfig{figure=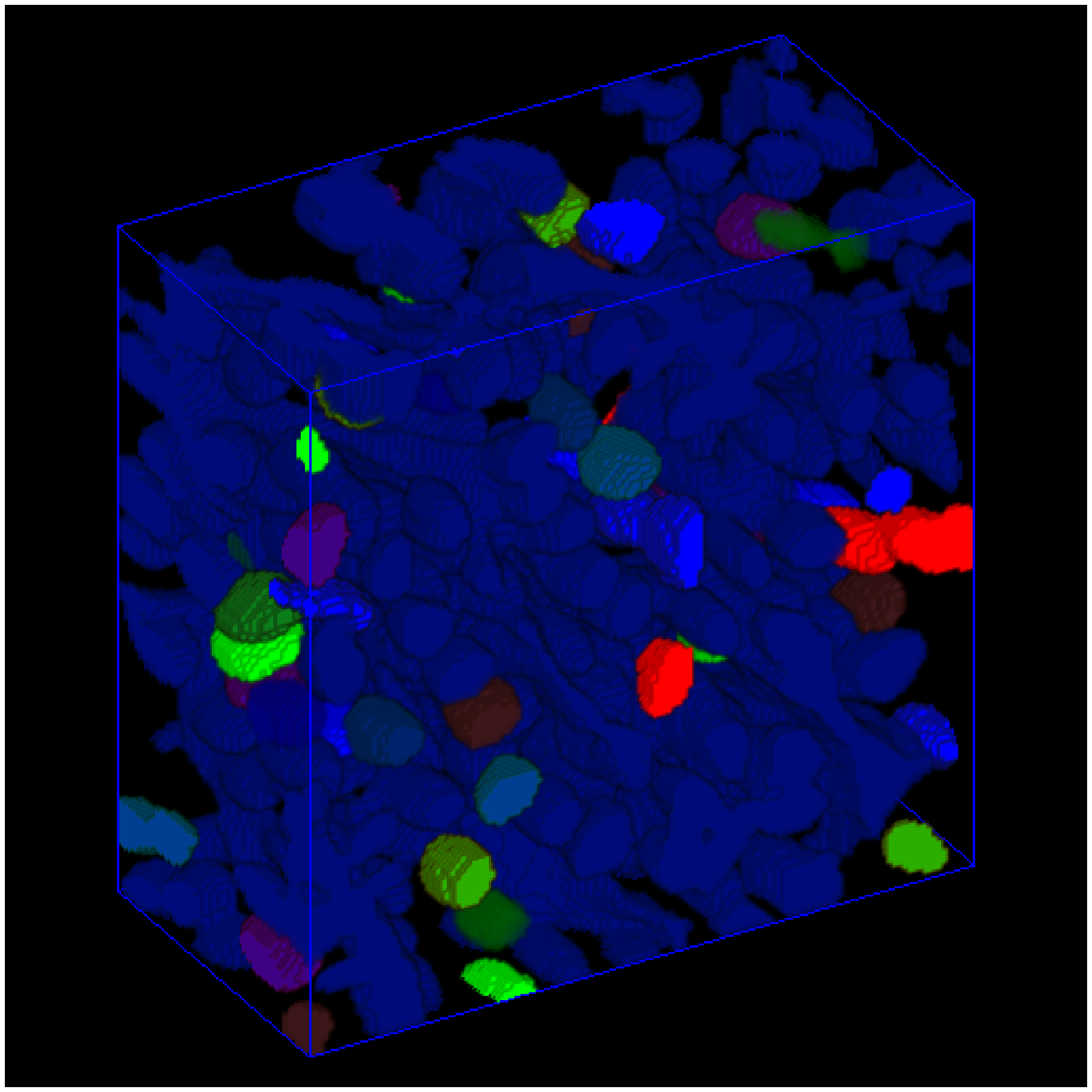,width=1.9cm}
}
\subfigure[]
{
	\epsfig{figure=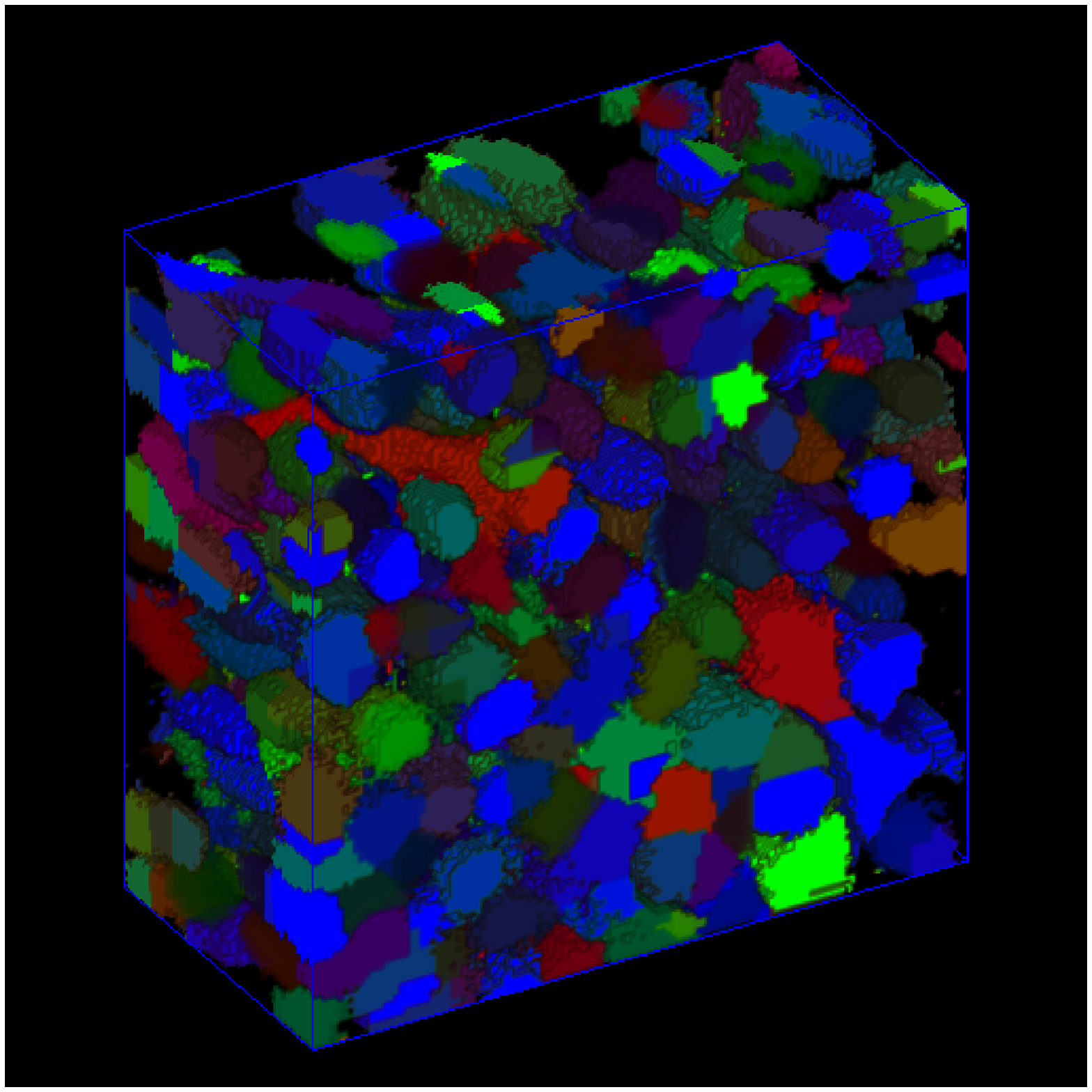,width=1.9cm}
}

\subfigure[]
{
	\epsfig{figure=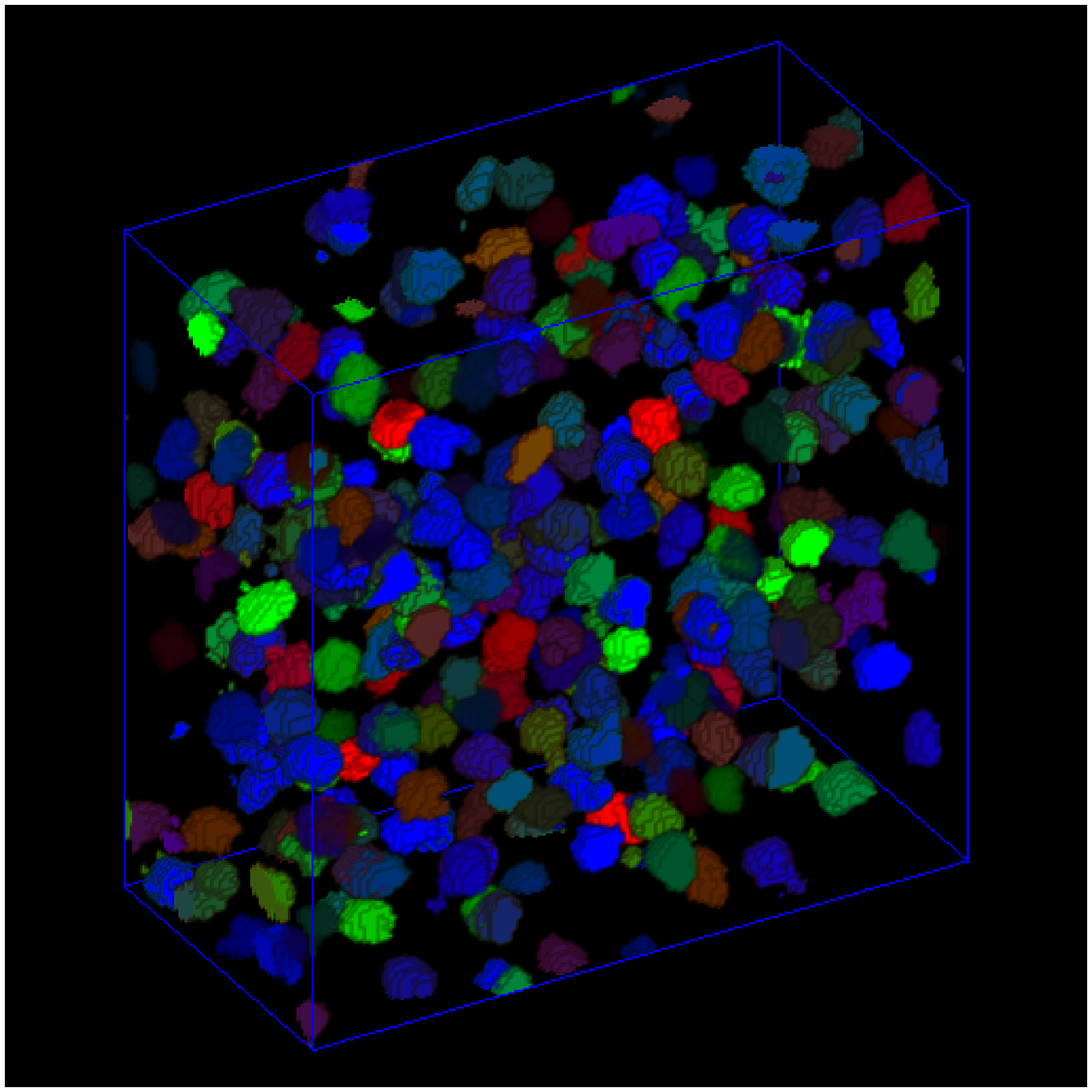,width=1.9cm}
}
\subfigure[]
{
	\epsfig{figure=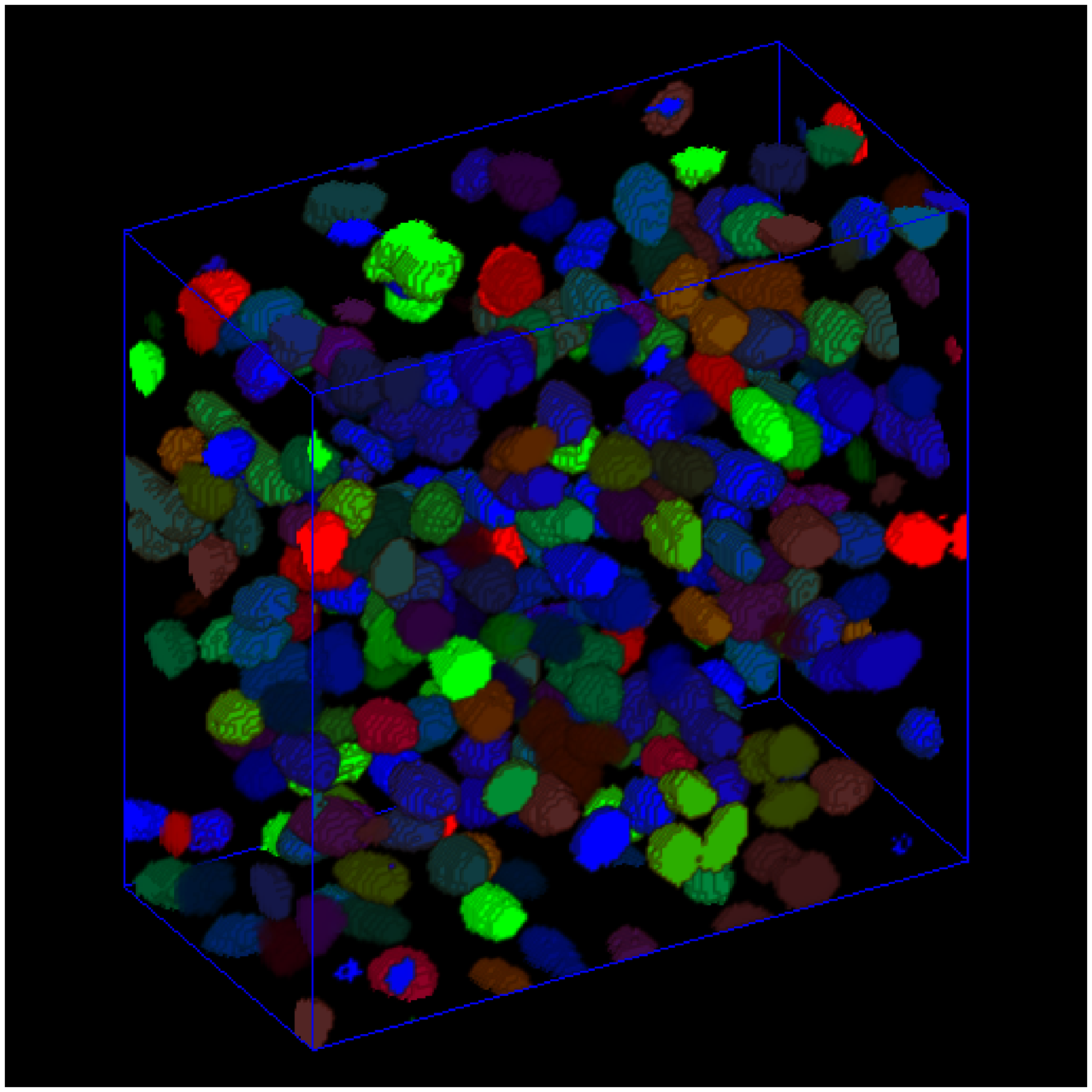,width=1.9cm}
}
\subfigure[]
{
	\epsfig{figure=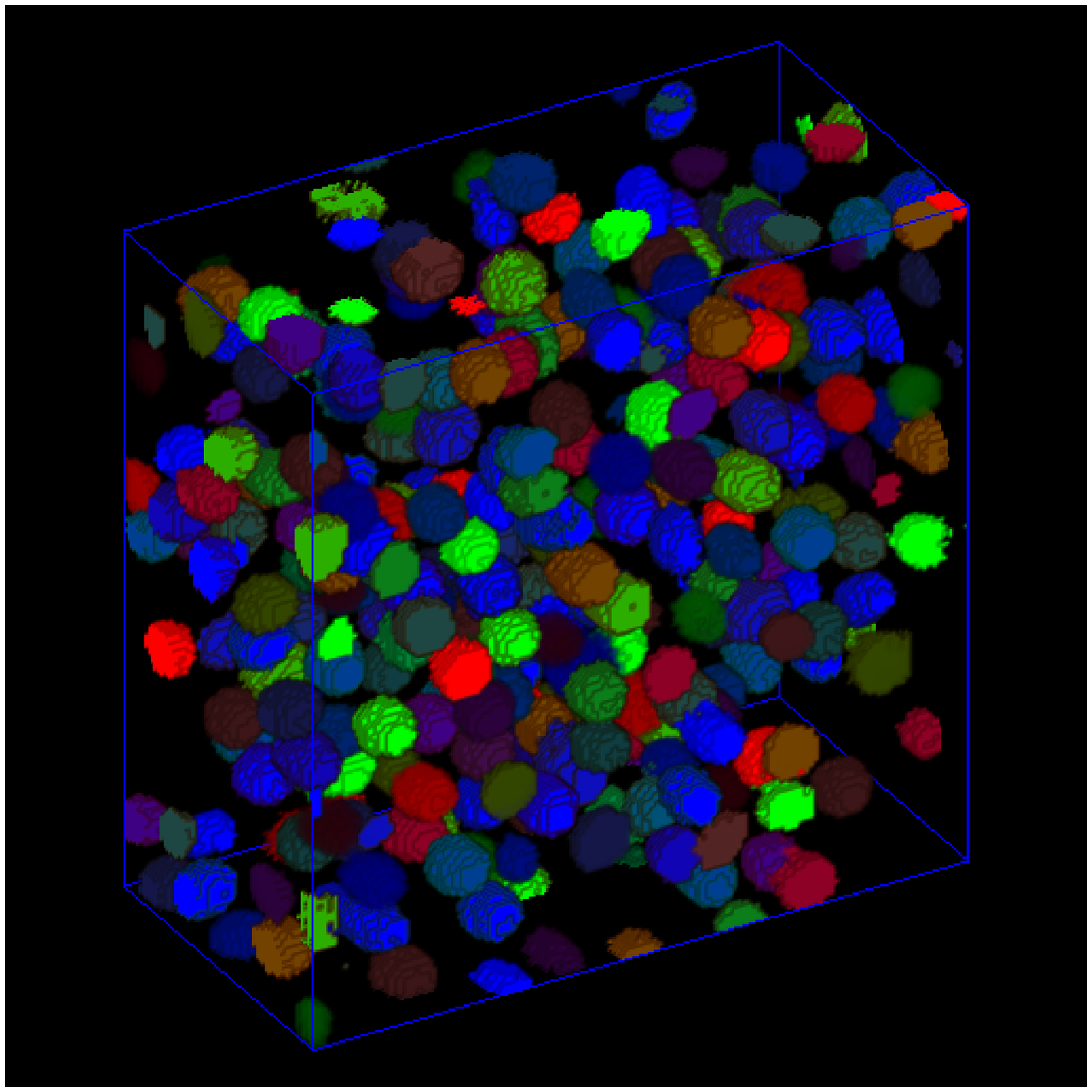,width=1.9cm}
}
\vspace{-0.15in}
\caption{Comparison of other segmentation methods and our proposed method of Data-I (a) original volume, (b) ground truth volume, (c) Squassh, (d) Watershed, (e) Purdue1, (f) Purdue2, (g) our proposed method}
\vspace{-0.15in}
\label{fig:comparison}
\end{figure}

Table \ref{tab:object} and Figure \ref{fig:comparison} show the object-based evaluation and the segmentation results visualized by Voxx \cite{clendenon2002} of other methods and our new proposed method for Data-I, respectively.
%Table \ref{tab:voxel} and Table \ref{tab:object} show the voxel-based evaluation and the object-based evaluation, respectively.
Squassh \cite{rizk2014} cannot distinguish nuclei and non-nuclei structures and cannot successfully separate touching objects.
For watershed \cite{vincent1991}, $I^{orig}$ is first binarized by a manually-selected threshold value of 64.
Thresholding cannot distinguish nuclei and non-nuclei structures and watershed technique over-segments foreground region.
Purdue1 can reject non-nuclei structures but still have a poor F1 score. 
Purdue2 can generate an accurate binary segmentation mask but cannot separate all touching nuclei.
%Purdue2 can generate an accurate binary segmentation mask.
%To separate touching nuclei, we used morphological operations with a 3D erosion, a 3D connected component for color-coding, and a 3D dilation with a sphere  of radius of 1  used as the structuring element.
%We observe that morphological operations cannot separate all touching nuclei.
Our proposed method, detecting the locations of nuclei and individually segmenting nuclei in 3D patches using the SpCycleGAN, can successfully segment and separate nuclei.

%\vspace{-0.15in}
%\begin{table}[htb]
%\centering
%{
%\caption{Voxel-wise evaluation for various methods for Data-I}
%\begin{tabular}{| c | c | c | c |}
	%\hline
	%& Accuracy & Type-I Error & Type-II Error\\
	%\hline
  %Watershed \cite{vincent1991} & 59.25\% & 40.75\% & 0.00\% \\
	%\hline
  %Squassh \cite{rizk2014} & 80.45\% & 19.54\% & 0.01\% \\
	%\hline
  %Purdue1 \cite{ho2018} & 93.65\% & 1.71\% & 4.64\% \\
	%\hline
	%Purdue2 \cite{fu2018} & 95.99\% & 2.32\% & 1.69\% \\
	%\hline
	%Proposed Method & 94.73\% & 3.85\% & 1.42\% \\
	%\hline
%\end{tabular}
%\vspace{-0.2in}
%\label{tab:voxel}
%}
%\end{table}

\vspace{-0.15in}
\begin{table}[htb]
\centering
{
\caption{Object-wise evaluation for various methods for Data-I}
\begin{tabular}{| c | c | c | c |}
	\hline
	& Precision & Recall & F1 score\\
	\hline
  Squassh \cite{rizk2014} & 85.07\% & 20.14\% & 32.57\% \\
	\hline
  Watershed \cite{vincent1991} & 51.14\% & 92.13\% & 65.78\% \\
	\hline
  Purdue1 \cite{ho2018} & 68.35\% & 90.22\% & 77.78\% \\
	\hline
	Purdue2 \cite{fu2018} & 91.20\% & 82.01\% & 86.36\% \\
	\hline
	Proposed Method & 93.47\% & 96.80\% & 95.10\% \\
	\hline
\end{tabular}
\vspace{-0.15in}
\label{tab:object}
}
\end{table}

%\vspace{-0.1in}

\vspace{-0.05in}
\section{Conclusions}
\vspace{-0.05in}
\label{sec:conclusions}
This paper presented a nuclei instance segmentation method using a center-extraction technique to detect the center locations of nuclei.
We individually segmented nuclei in 3D patches surrounding the nuclei.
Our method can successfully segment nuclei visually and numerically.
In the future we plan to develop a synthetic volume generation model which can produce synthetic nuclei with other shapes.
\vspace{-0.05in}
\section*{Acknowledgment}
\vspace{-0.05in}
Data-I was provided by Malgorzata Kamocka of Indiana University and was collected at the Indiana Center for Biological Microscopy. 
Data-II was provided by Sherry Clendenon collected while at the Indiana Center for Biological Microscopy. 
She is currently at the Department of Intelligent Systems Engineering of Indiana University.
\bibliographystyle{IEEEbib}
\bibliography{refs}

\begin{thebibliography}{10}

\bibitem{dunn2002}
K.W. Dunn, R.M. Sandoval, K.J. Kelly, P.C. Dagher, G.A. Tanner, S.J. Atkinson,
  R.L. Bacallao, and B.A. Molitoris,
\newblock ``Functional studies of the kidney of living animals using multicolor
  two-photon microscopy,''
\newblock {\em American Journal of Physiology-Cell Physiology}, vol. 283, no.
  3, pp. C905--C916, September 2002.

\bibitem{vonesch2006}
C.~Vonesch, F.~Aguet, J.~Vonesch, and M.~Unser,
\newblock ``The colored revolution of bioimaging,''
\newblock {\em IEEE Signal Processing Magazine}, vol. 23, no. 3, pp. 20--31,
  May 2006.

\bibitem{vincent1991}
L.~Vincent and P.~Soille,
\newblock ``Watersheds in digital spaces: an efficient algorithm based on
  immersion simulations,''
\newblock {\em IEEE Transactions on Pattern Analysis and Machine Intelligence},
  vol. 13, no. 6, pp. 583--598, June 1991.

\bibitem{yang2006}
X.~Yang, H.~Li, and X.~Zhou,
\newblock ``Nuclei segmentation using marker-controlled watershed, tracking
  using mean-shift, and {Kalman} filter in time-lapse microscopy,''
\newblock {\em IEEE Transactions on Circuits and Systems I: Regular Papers},
  vol. 53, no. 11, pp. 2405--2414, November 2006.

\bibitem{delgado2015}
R.~Delgado-Gonzalo, V.~Uhlmann, D.~Schmitter, and M.~Unser,
\newblock ``Snakes on a plane: A perfect snap for bioimage analysis,''
\newblock {\em IEEE Signal Processing Magazine}, vol. 32, no. 1, pp. 41--48,
  January 2015.

\bibitem{dufour2005}
A.~Dufour, V.~Shinin, S.~Tajbakhsh, N.~Guillen-Aghion, J.C. Olivo-Marin, and
  C.~Zimmer,
\newblock ``Segmenting and tracking fluorescent cells in dynamic {3-D}
  microscopy with coupled active surfaces,''
\newblock {\em IEEE Transactions on Image Processing}, vol. 14, no. 9, pp.
  1396--1410, September 2005.

\bibitem{rizk2014}
A.~Rizk, G.~Paul, P.~Incardona, M.~Bugarski, M.~Mansouri, A.~Niemann,
  U.~Ziegler, P.~Berger, and I.F. Sbalzarini,
\newblock ``Segmentation and quantification of subcellular structures in
  fluorescence microscopy images using {Squassh},''
\newblock {\em Nature Protocols}, vol. 9, no. 3, pp. 586--596, February 2014.

\bibitem{lecun2015}
Y.~LeCun, Y.~Bengio, and G.~Hinton,
\newblock ``Deep learning,''
\newblock {\em Nature}, vol. 521, pp. 436--444, May 2015.

\bibitem{litjens2017}
G.~Litjens, T.~Kooi, B.E. Bejnordi, A.A.A. Setio, F.~Ciompi, M.~Ghafoorian,
  J.A.W.M. van~der Laak, B.~van Ginneken, and C.I. Sanchez,
\newblock ``A survey on deep learning in medical image analysis,''
\newblock {\em Medical Image Analysis}, vol. 42, pp. 60--88, July 2017.

\bibitem{chen2016}
H.~Chen, X.~Qi, L.~Yu, and P.-A. Heng,
\newblock ``{DCAN}: Deep contour-aware networks for accurate gland
  segmentation,''
\newblock {\em Proceedings of the IEEE Conference on Computer Vision and
  Pattern Recognition}, pp. 2487--2496, June 2016,
\newblock {Las Vegas, NV}.

\bibitem{graham2018}
S.~Graham and N.M. Rajpoot,
\newblock ``{SAMS-NET}: Stain-aware multi-scale network for instance-based
  nuclei segmentation in histology images,''
\newblock {\em Proceedings of the IEEE International Symposium on Biomedical
  Imaging}, pp. 590--594, April 2018,
\newblock {Washington, D.C.}

\bibitem{falk2019}
T.~Falk, D.~Mai, R.~Bensch, O.~Cicek, A.~Abdulkadir, Y.~Marrakchi, A.~Bohm,
  J.~Deubner, Z.~Jackel, K.~Seiwald, A.~Dovzhenko, O.~Tietz, C.~Dal Bosco,
  S.~Walsh, D.~Saltukoglu, T.L. Tay, M.~Prinz, K.~Palme, M.~Simons, I.~Diester,
  T.~Brox, and O.~Ronneberger,
\newblock ``{U-Net}: deep learning for cell counting, detection, and
  morphometry,''
\newblock {\em Nature Method}, vol. 16, pp. 67--70, January 2019.

\bibitem{cicek2016}
O.~Cicek, A.~Abdulkadir, S.S. Lienkamp, T.~Brox, and O.~Ronneberger,
\newblock ``{3D U-Net: Learning dense volumetric segmentation from sparse
  annotation},''
\newblock {\em Proceedings of the Medical Image Computing and Computer-Assisted
  Intervention}, pp. 424--432, October 2016,
\newblock {Athens, Greece}.

\bibitem{ho2017b}
D.J. Ho, C.~Fu, P.~Salama, K.W. Dunn, and E.J. Delp,
\newblock ``Nuclei segmentation of fluorescence microscopy images using three
  dimensional convolutional neural networks,''
\newblock {\em Proceedings of the Computer Vision for Microscopy Image Analysis
  workshop at Computer Vision and Pattern Recognition}, pp. 834--842, July
  2017,
\newblock {Honolulu, HI}.

\bibitem{ho2018}
D.J. Ho, C.~Fu, P.~Salama, K.W. Dunn, and E.J. Delp,
\newblock ``Nuclei detection and segmentation of fluorescence microscopy images
  using three dimensional convolutional neural networks,''
\newblock {\em Proceedings of the IEEE International Symposium on Biomedical
  Imaging}, pp. 418--422, April 2018,
\newblock {Washington, D.C.}

\bibitem{goodfellow2014}
I.~Goodfellow, J.~Pouget-Abadie, M.~Mirza, B.~Xu, D.~Warde-Farley, S.~Ozair,
  A.~Courville, and Y.~Bengio,
\newblock ``Generative adversarial nets,''
\newblock {\em Proceedings of the Advances in Neural Information Processing
  Systems}, pp. 2672--2680, December 2014,
\newblock {Montreal, Canada}.

\bibitem{zhu2017}
J.Y. Zhu, T.~Park, P.~Isola, and A.A. Efros,
\newblock ``Unpaired image-to-image translation using cycle-consistent
  adversarial networks,''
\newblock {\em Proceedings of the IEEE International Conference on Computer
  Vision}, pp. 2223--2232, October 2017,
\newblock {Venice, Italy}.

\bibitem{fu2018}
C.~Fu, S.~Lee, D.J. Ho, S.~Han, P.~Salama, K.W. Dunn, and E.J. Delp,
\newblock ``Three dimensional fluorescence microscopy image synthesis and
  segmentation,''
\newblock {\em Proceedings of the Computer Vision for Microscopy Image Analysis
  workshop at Computer Vision and Pattern Recognition}, pp. 2302--2310, June
  2018,
\newblock {Salt Lake City, UT}.

\bibitem{paszke2017}
A.~Paszke, S.~Gross, S.~Chintala, G.~Chanan, E.~Yang, Z.~DeVito, Z.~Lin,
  A.~Desmaison, L.~Antiga, and A.~Lerer,
\newblock ``Automatic differentiation in pytorch,''
\newblock {\em Proceedings of the Autodiff Workshop at the Advances in Neural
  Information Processing Systems}, pp. 1--4, December 2017,
\newblock {Long Beach, CA}.

\bibitem{clendenon2002}
J.L. Clendenon, C.L. Phillips, R.M. Sandoval, S.~Fang, and K.W. Dunn,
\newblock ``Voxx: a {PC}-based, near real-time volume rendering system for
  biological microscopy,''
\newblock {\em American Journal of Physiology-Cell Physiology}, vol. 282, no.
  1, pp. C213--C218, January 2002.

\bibitem{kingma2014}
D.P. Kingma and J.~Ba,
\newblock ``Adam: A method for stochastic optimization,''
\newblock {\em arXiv preprint arXiv:1412.6980}, pp. 1--15, December 2014.

\bibitem{yu2016}
F.~Yu and V.~Koltun,
\newblock ``Multi-scale context aggregation by dilated convolutions,''
\newblock {\em arXiv preprint arXiv:1511.07122}, pp. 1--13, April 2016.

\bibitem{powers2011}
D.M.W. Powers,
\newblock ``Evaluation: from {Precision, Recall and F-measure to ROC,
  Informedness, Markedness and Correlation},''
\newblock {\em Journal of Machine Learning Technologies}, vol. 2, no. 1, pp.
  37--63, December 2011.

\bibitem{yushkevich2006}
P.A. Yushkevich, J.~Piven, H.C. Hazlett, R.G. Smith, S.~Ho, J.C. Gee, and
  G.~Gerig,
\newblock ``User-guided {3D} active contour segmentation of anatomical
  structures: Significantly improved efficiency and reliability,''
\newblock {\em NeuroImage}, vol. 31, no. 3, pp. 1116--1128, July 2006.

\end{thebibliography}

\end{document}